\documentclass[10pt]{article}

\usepackage{amsmath, amssymb,array}

\newtheorem{thm}{\bf Theorem}[section]
\newtheorem{lem}[thm]{\bf Lemma}        %% lemmas, props, cor, etc
\newtheorem{prop}[thm]{\bf Proposition}  %% with the theorems.
\newtheorem{cor}[thm]{\bf Corollary}
\newtheorem{deff}[thm]{\bf Definition}

\newtheorem{expl}[thm]{\bf Example}       
\newtheorem{remark}[thm]{\bf Remark}

\def\Bbox{
{\unskip\nobreak\hfil\penalty50
\hskip1em\hbox{}\nobreak\hfil{\lower .5pt \hbox{$\Box$}}
\parfillskip=0pt \finalhyphendemerits=0 \par}
}

\def\eop{
\ifmmode {\hbox{\Bbox}} \else \Bbox \fi
}

\def\bbox{
\ifmmode {\hbox{\bbox}} \else \Bbox \fi
}

\newtheorem{example}[thm]{\bf Example}

\def\N{\mathbb{N}}
\def\B{\mathbb{B}}
\def\R{\mathbb{R}}
\def\oR{\overline{\mathbb{R}}}

\def\llangle{{\langle {\kern -.2em} \langle}}
\def\rrangle{{\rangle {\kern -.2em} \rangle}}
\def\val{\mathrm{val}}
\def\os{{\,\oplus\,}}
\def\rat{{\mathsf{rat}}}
\def\cA{{\mathcal{A}}}
\def\avg{\mathrm{avg}}
\def\disc{\mathrm{disc}}
\def\supp{\mathrm{supp}}
\def\wt{{\mathrm{wt}}}

\begin{document}

\title{{\bf Conway and iteration hemirings \\
Parts 1 and 2}}

\author{M. Droste\\
{\small Dept. of Computer Science}\\
{\small University of Leipzig}\\
{\small Germany}
\and 
Z. \'Esik\thanks{Partially supported by the European Union and
co-funded by the European Social Fund.
Project title: Telemedicine-focused research activities in the field of
Mathematics, Informatics and Medical sciences.
Project number: TÁMOP-4.2.2.A-11/1/KONV-2012-0073.}\\
{\small Dept. of Computer Science}\\
{\small University of Szeged}\\
{\small Hungary}\\
\and
W. Kuich\\
{\small Inst. Discrete Mathematics and Geometry}\\
{\small Technical University of Vienna}\\
{\small Austria}
}

\date{\empty}

\maketitle

\begin{abstract}
Conway hemirings are Conway semirings without a multiplicative 
unit. We also define iteration hemirings as Conway hemirings 
satisfying certain identities associated with the finite groups. 
Iteration hemirings are iteration semirings without a multiplicative 
unit. We provide an analysis of the relationship
between Conway hemirings and (partial) Conway semi\-rings 
and describe several free constructions.
In the second part of the paper we define and study hemimodules 
of Conway and iteration hemirings, and show their applicability 
in the analysis of quantitative aspects 
of the infinitary behavior of weighted transition systems. These include discounted
and average computations of weights investigated recently in \cite{Chatterjeeetal, Chatterjeeetal2}.
\end{abstract}

\begin{center} 
{\bf \Large Part 1}
\end{center}

\vspace*{.2in}

\section{Introduction to Part 1}

A Conway semiring \cite{BEbook,Conway} is a semiring $S$ equipped with a star operation 
$^* : S \to S$, satisfying two important identities of regular languages, 
the `sum star' and `product star' identities. The main interest 
in Conway semirings in computer science is due to the fact that 
it is possible to define automata \cite{BEbook} in Conway semirings as 
a generalization of the classic notion of nondeterministic and
weighted finite automata, and to develop (a part of) 
the theory of finite automata in an axiomatic setting, including a general Kleene theorem. 
Conway semiring-semimodule pairs $(S,V)$ are equipped with 
both a star operation $^*: S \to S$ and an omega operation 
$^\omega: S \to V$. These structures were introduced in \cite{BEbook} with the intention to 
serve as an axiomatic framework for automata on $\omega$-words. 
However, in several natural models
such as rational series over words and/or $\omega$-words 
with coefficients in certain semirings, the star and omega power 
operations cannot be defined for all elements. 
This leads to the question of validity of weighted automata studied 
extensively in the recent paper \cite{SacLomb}.
There are two simple possible ways to solve this problem. The first
solution amounts to introducing partial Conway semirings and 
partial Conway semiring-semimodule pairs, cf. \cite{BET,Esikpartial}, 
where the domain of the star and omega operations is restricted to certain 
ideal elements, typically not including the multiplicative unit of the 
semiring. The disadvantage of this approach is that star and 
omega become partial operations. 
The second solution is to simply discard all those elements,
including the multiplicative unit, for which it is not possible 
to define the star and/or omega operation in a reasonable way.
 
This second approach leads to Conway hemirings $H$ \cite{DrosteKuich},
equipped with a total plus operation $^+: H \to H$. In this paper, 
we also define Conway hemiring-hemimodule pairs $(H,V)$, which,
in addition to the plus operation, also possess a total omega 
operation $^\omega: H \to V$. 
We offer a study of Conway hemirings and Conway hemiring-hemimodule pairs 
in relation to (partial) Conway semirings and Conway semiring-semimodule pairs.  
In Theorem~\ref{thm-cop1}, we show how to add freely (at least under 
certain conditions) a semiring $S_0$ to a Conway hemiring $H$
to obtain a partial Conway semiring $S_0 \os H$. Then, in Theorem~\ref{thm-free1}, 
we use this result to describe the free partial Conway semiring generated by a Conway hemiring,
and in Theorem~\ref{thm-cop2}, we show that when $S_0$ is a Conway semiring, then $S_0 \os H$ 
may also be turned into a Conway semiring with a totally defined star operation 
in a canonical way. The proof of this latter fact uses Theorem~\ref{thm-ext} (an improvement of
the Matrix Extension Theorem \cite{BEbook}), which is of independent interest. 
We also define iteration hemirings as Conway hemirings satisfying a variant
of Conway's group identities \cite{Conway} associated with the finite groups,
and study their relation to iteration semirings. We prove that when $S_0$ 
is an iteration semiring and $H$ is an iteration hemiring, then $S_0 \oplus H$ 
is an iteration semiring (Theorem~\ref{thm-cop22}). The importance of the notion 
of iteration hemirings is shown by the fact that for every alphabet $A$,
the hemiring $N^\rat\llangle A^+\rrangle$ of rational power series of nonempty 
words with coefficients in the semiring $\N$ of natural numbers is the free 
iteration hemiring, freely generated by $A$ (Theorem~\ref{thm-free iteration hemiring}). 

In the second part of the paper, we define Conway and iteration hemiring-hemimodule pairs 
and apply them in the analysis of quantitative aspects 
of the long time behavior of transition systems, as introduced and investigated recently in \cite{Chatterjeeetal, Chatterjeeetal2}.

\section{Conway hemirings}

We assume familiarity with basic concepts of \emph{semirings} $S = (S,+.\cdot,0,1)$
as defined in \cite{Golan,KuichSalomaa}. Examples of semirings include the semiring of 
natural numbers $\N$ and the boolean semiring $\B$. A semiring $S$ is
\emph{idempotent} is $x + x = x$ for all $x \in S$. For example, $\B$ is 
idempotent. 

Recall from \cite{BEbook} that a \emph{Conway semiring} is a semiring 
$S= (S,+,\cdot,0,1)$, equipped with a star operation 
$^*: S \to S$, satisfying the identities:
\newcolumntype{L}[1]{>{\raggedright\let\newline\\\arraybackslash\hspace{0pt}}m{#1}}
\newcolumntype{C}[1]{>{\centering\let\newline\\\arraybackslash\hspace{0pt}}m{#1}}
\newcolumntype{R}[1]{>{\raggedleft\let\newline\\\arraybackslash\hspace{0pt}}m{#1}}
\begin{flalign}
\label{eq-product star}
& \begin{array}{@{}L{7cm}R{2cm}C{0.5cm}L{3cm}}
\text{\it sum star identity} & $(x+y)^*$ & $=$ & $(x^*y)^*x^*$
\end{array} & \\
\label{eq-sum star}
& \begin{array}{@{}L{7cm}R{2cm}C{0.5cm}L{3cm}}
\text{\it product star identity} & $(xy)^*$ & $=$ & $1 + x(yx)^*y$
\end{array} &
\end{flalign}
for all $x,y \in S$. 
It is known that the following identities hold in
every Conway semiring:
\begin{flalign}
& \begin{array}{@{}L{7cm}R{2cm}C{0.5cm}L{3cm}}
\text{\it sum star identity} & $(x+y)^*$ & $=$ & $x^* (yx^*)^*$
\end{array} & \\
& \begin{array}{@{}L{7cm}R{2cm}C{0.5cm}L{3cm}}
\text{\it simplified product star identity} & $(xy)^* x$ & $=$ & $x (yx)^*$
\end{array} & \\
& \begin{array}{@{}L{7cm}R{2cm}C{0.5cm}L{3cm}}
\text{\it fixed point identity} & $xx^* + 1$ & $=$ & $x^*$
\end{array} & \\
& \begin{array}{@{}L{7cm}R{2cm}C{0.5cm}L{3cm}}
\text{\it dual fixed point identity} & $x^*x + 1$ & $=$ & $x^*$
\end{array} & \\
& \begin{array}{@{}L{7cm}R{2cm}C{0.5cm}L{3cm}}
\text{\it unary product star identity} & $xx^*$ & $=$ & $x^*x$
\end{array} & \\
& \begin{array}{@{}L{7cm}R{2cm}C{0.5cm}L{3cm}}
\text{\it zero star identity} & $0^*$ & $=$ & $1$
\end{array} &
\end{flalign}

A morphism of Conway semirings is a semiring morphism preserving the star operation.
Conway semirings are implicit in \cite{Conway}.

In a Conway semiring $S$, we may define a \emph{plus operation}
$x\mapsto x^+$ by $x^+ = xx^* = x^*x$.
This operation satisfies following identities:
\begin{flalign}
& \begin{array}{@{}L{7cm}R{2cm}C{0.5cm}L{3cm}}
\text{\it sum plus identity} & $(x+y)^+$ &$=$& $(x^*y)^+x^* + x^+$
\end{array} & \\
& \begin{array}{@{}L{7cm}R{2cm}C{0.5cm}L{3cm}}
\text{\it simplified product plus identity} & $(xy)^+x$ &$=$& $x(yx)^+$
\end{array} & \\
& \begin{array}{@{}L{7cm}R{2cm}C{0.5cm}L{3cm}}
\text{\it plus fixed point identity} & $xx^+ + x$ &$=$& $x^+$
\end{array} & \\
& \begin{array}{@{}L{7cm}R{2cm}C{0.5cm}L{3cm}}
\text{\it dual plus fixed point identity} & $x^+x + x$ &$=$& $x^+$
\end{array} & \\
& \begin{array}{@{}L{7cm}R{2cm}C{0.5cm}L{3cm}}
\text{\it unary product plus identity} & $x^+x$ &$=$& $xx^+$
\end{array} & \\
& \begin{array}{@{}L{7cm}R{2cm}C{0.5cm}L{3cm}}
\text{\it zero plus identity} & $0^+$ &$=$& $0.$
\end{array} & 
\end{flalign}

Recall from \cite{Golan} that a hemiring is defined as a semiring but without requiring a multiplicative unit. 
Clearly, every semiring is a hemiring. 
 In \cite{DrosteKuich}, a \emph{Conway hemiring} is defined as a hemiring $H$ 
 equipped with a \emph{plus operation} satisfying the sum plus, simplified product plus, 
unary product plus and plus fixed point identities. Morphisms of 
Conway hemirings are hemiring morphisms preserving the plus operation.

The following fact was noted in \cite{DrosteKuich}.

\begin{lem}
\label{lem-first}
A semiring $S$ equipped with a star operation $^*: S \to S$ 
is a Conway semiring iff it is a Conway hemiring with the plus 
operation $^+: S \to S$, $s\mapsto s^+ = ss^*$. Moreover, a semiring 
$S$, equipped with a plus operation $^+: S \to S$, is a Conway hemiring 
iff $S$ is a Conway semiring with the star operation 
$^*: S \to S$ defined by $s^* = 1 + s^+$.
\end{lem}
In short, a Conway semiring is a Conway hemiring which is a semiring.
For any $x,y$ in a Conway hemiring $H$, we will write $x^*y$ for $x^+y + x$ and 
$yx^*$ for $y + yx^+$. When $H$ is a Conway hemiring which is a semiring, 
we also define $x^* = 1 + x^+$ for all $x \in H$, and call this operation
the star operation determined by the plus operation.

It is also clear that a semiring morphism between Conway semirings 
is a Conway semiring morphism iff it is Conway hemiring morphism.

\section{Extending a Conway hemiring with a semiring}
\label{sec-extend}

An \emph{ideal} of a semiring $S$ is a set $I \subseteq S$
containing $0$ which is closed under the sum operation 
and left and right multiplication with any element of $S$, 
i.e., $0 \in I$, $I + I \subseteq I$ and $SI \cup IS \subseteq I$.
Following \cite{BET}, a \emph{partial Conway semiring}
$(S,I,^*)$ consists of a semiring $S$, a distinguished 
ideal $I$ of $S$, equipped with a star operation 
$^*: I \to S$, satisfying the star sum identity (\ref{eq-sum star}) 
for all $x,y \in I$ and the product star identity 
(\ref{eq-product star}) for all $x,y\in S$ such that  
$x$ or $y$ is in $I$. A morphism of partial Conway semirings
is a semiring morphism that preserves the distinguished ideal
and the star operation. 

When $(S,I,^*)$ is a partial Conway semiring, the plus
operation determined by the star operation 
$x \mapsto x^+ = xx^* = x^*x$ maps $I$ into itself,
and it is not difficult to see that $I$, equipped
with this plus operation, is a Conway hemiring. A certain converse
of this fact was proved in \cite{DrosteKuich}. In this 
section we provide a generalization of this result by showing 
that, under a natural condition, 
for every Conway hemiring $H$ and semiring $S_0$, 
there is a partial Conway semiring $(S,H,^*)$ containing $S_0$ as 
a subsemiring, with distinguished ideal $H$ such that 
the star operation is the one determined by the plus 
operation on $H$.  

A \emph{left action} of a semiring $S$ on a hemiring $H$ 
is a function $S \times H \to H$, $(x,a) \mapsto xa$, subject 
to the usual laws:
\begin{eqnarray*}
(x+y)a &=& xa+ ya\\
(xy)a &=& x(ya)\\
0a &=& 0\\
1a &=& a\\
x(a+b) &=& xa + xb\\
x(ab) &=& (xa)b\\
x0 &=& 0,
\end{eqnarray*}
for all $x,y \in S$ and $a,b\in H$. A right action of $S$ on $H$ is defined similarly. 
A \emph{bi-action} is both a left action and a right action, which additionally satisfies 
\begin{eqnarray*}
(xa)y &=& x(ay),
\end{eqnarray*}
for all $x,y \in S$ and $a \in A$.

Suppose now that $H$ is a Conway hemiring and $S_0$ is a semiring
with a bi-action on $H$ such that
\begin{eqnarray}
\label{eq-action}
(xa)^+x = x(ax)^+ \quad {\rm and}\quad (ax)^+a = a(xa)^+
\end{eqnarray}
 hold for all $x\in S_0$ and $a \in H$.
Define $S = S_0 \times H$ as the Cartesian product of
$S_0$ and $H$, equipped with pointwise sum operation 
and the product 
$$(x,a)(y,b) = (xy ,xb + ay + ab).$$
It is easy to see that $S$ is in fact a semiring with $0 = (0,0)$ 
and $1 = (1,0)$. Moreover, 
$\{0\} \times H$ is an ideal of $S_0 \times H$. We define a star operation 
$^*: \{0\}\times H \to S$ by 
 % $$s^* = (x^*,  (x^*a)^+ x^*)$$
 $$s^* = (1,a^+),$$
for all $s = (0,a) \in S$. 

Alternatively, we may think of $S$ as the set of all \emph{formal} sums 
$x + a$ with $x\in S_0$ and $a \in H$. 
The sum and product operations are 
\begin{eqnarray*}
 (x+a) + (y +b) &=& (x+y) +(a + b)\\
 (x+a)(y+b) &=& xy +( xb + ay + ab),
\end{eqnarray*}
and the constants are given by $0 = 0 + 0$ and $1 = 1+0$. 
(Note that multiplication extends the action.)
The star operation is $(0+a)^* = 1 + a^+$, for all $a \in H$.

Clearly, $S_0$ embeds in $S$ by the semiring morphism
$\kappa: x \mapsto (x,0)$ (or $\kappa: x \mapsto x+0$), and $H$ embeds in $S$ 
by the Conway hemiring morphism $\lambda: a \mapsto (0,a)$ (or $\lambda:a \mapsto 0+a$).
Below we will mainly use additive notation and identify $x\in S_0$ with 
$x\kappa$ and $a\in H$ with $a\lambda$. In particular, we view 
$S_0$ as a subsemiring and $H$ as an ideal of $S$. 
Using this identification, we have that $a^* = 1 + a^+$ for all $a\in H$.

\begin{prop}
\label{prop-ext}
Under the above assumptions, $(S,H,^*)$ is a partial Conway semiring. 
\end{prop}
 
{\sl Proof.} 
Easy calculations show that $S$ is a semiring and $H$ is an ideal of $S$.
The sum plus, plus fixed point and unary product plus 
identities clearly hold.
To complete the proof, we need to show that the simplified product plus
identity holds as well. Suppose that $s\in S$ and $a\in H$.
We want to prove that $(sa)^+s = s(as)^+$ and $(as)^+a = 
a(sa)^+$. When $s = y \in S_0$, these equalities hold by 
assumption. Suppose now that $s = y + b$ with $y\in S_0$ and
$b \in H$. 

We have 
\begin{eqnarray*}
(as)^+a 
&=& (a(y+b))^+a\\
&=& (ay + ab)^+ a\\
&=& ((ay)^*ab)^+(ay)^*a + (ay)^+a\\
&=& ( a(ya)^*b)^+a(ya)^* + a(ya)^+\\
&=& a((ya)^*ba)^+(ya)^*+ a(ya)^+\\
&=& a(ya + ba)^+\\
&=& a(sa)^+.
\end{eqnarray*}
Also,
\begin{eqnarray*}
(sa)^+s
&=& ((y+b)a)^+(y + b)\\
&=& (ya +ba)^+ (y + b) \\
&=& (ya + ba)^+y + (ya + ba)^+b\\
&=&  ya(ya + ba)^*y + ba(ya + ba)^*y + ya(ya + ba)^*b + ba(ya + ba)^*b\\
&=& \alpha_1 + \beta_1 + \gamma_1 + \delta_1.
\end{eqnarray*}
and
\begin{eqnarray*}
s(as)^+ &=& (y+b)(a(y+b))^+\\
&=& (y+b)(ay + ab)^+\\
&=& y(ay + ab)^+ + b(ay + ab)^+\\
&=& y(ay + ab)^*ay + y(ay + ab)^*ab + b(ay + ab)^+ay + b(ay + ab)^*ab\\
&=& \alpha_2 + \gamma_2 +\beta_2 + \delta_2.
\end{eqnarray*}
Now 
\begin{eqnarray*}
(ay + ab)^*a &=& ((ay)^*ab)^*(ay)^*a\\
&=& (a(ya)^*b)^*a(ya)^*\\
&=& a((ya)^*ba)^*(ya)^*\\
&=& a(ya + ba)^*,
\end{eqnarray*}
and it follows that $\alpha_1 = \alpha_2,\ldots,\delta_1 =\delta_2$
and $(sa)^+s = s(as)^+$. 
Apply now Lemma~\ref{lem-first}. \eop 

Note that in order that $(S,H,^*)$ be a partial Conway semiring, 
it is necessary to have (\ref{eq-action}).
We will denote $S$ by $S_0 \oplus H$ and call it the 
\emph{extension of $H$ by $S_0$}.

\begin{thm}
\label{thm-cop1}
Suppose that $H$ is a Conway hemiring, $S_0$ is a semiring
with a bi-action on $H$ such that 
(\ref{eq-action}) holds for all $x \in S_0$ 
and $a \in H$. Suppose that $(S',I',^*)$ is a partial Conway semiring, 
$\varphi: S_0 \to S'$ is a semiring morphism and $\psi: 
H \to I'$ is a `compatible' Conway hemiring morphism with
\begin{eqnarray}
\label{eq-compatible1}
(x\varphi)(a\psi) &=& (xa)\psi\\
\label{eq-compatible2}
(a\psi)(x\varphi) &=& (ax)\varphi,
\end{eqnarray}
for all $x \in S_0$ and $a \in A$.
Then there is a unique partial Conway semiring morphism
$\tau: S = S_0 \oplus H \to S'$ such that $\kappa\tau = \varphi$ and 
$\lambda\tau = \psi$, i.e., such that $\tau$ extends both $\varphi$ and $\psi$.
\end{thm}

{\sl Proof.} We already know that $(S,H,^*)$ is a partial Conway semiring.
Given $\varphi$ and $\psi$, we define $s\tau = x\varphi + a\psi$ 
for all $s = x + a \in S$. We need to show that $\tau$ is a morphism of
partial Conway semirings extending $\varphi$ and $\psi$. Throughout
the proof, $s_1 = x + a$ and $s_2 = y+b$ are in $S$ with 
$x,y\in S_0$ and $a,b\in H$.

It is clear that $\tau$ extends $\varphi$ and $\psi$, 
i.e., $\kappa\tau = \varphi$ and $\lambda \tau = \psi$.
Also, $\tau$ preserves $0$ and $1$ and the distinguished ideal.

{\em Claim: $\tau$ preserves sum.} Indeed, we have 
\begin{eqnarray*}
(s_1 + s_2)\tau &=&((x+y) + (a+b))\tau\\
&=& (x+y)\varphi + (a+b)\psi\\
&=& x\varphi + y \varphi + a\psi + b\psi \\
&=& (x\varphi + a\psi) + (y\varphi + b\psi)\\
&=&s_1\tau + s_2\tau.
\end{eqnarray*}

{\em Claim: $\tau$ preserves product.} Indeed,
\begin{eqnarray*}
(s_1s_2)\tau &=& ((xy +(xb + ay + ab))\tau\\
&=&(xy)\varphi + (xb + ay + ab)\psi\\
&=& (x\varphi)(y \varphi) + (xb)\psi + (ay)\psi + (ab)\psi\\
&=& (x\varphi)(y \varphi) + (x\varphi)(b\psi) + (a\psi)(y\varphi) + (a\psi)(b\psi)\\
&=& (x\varphi + a\psi)(y\varphi + b\psi)\\
&=&  (s_1\tau)(s_2\tau).
\end{eqnarray*}

{\em Claim: $\tau$ preserves $^*$.} Indeed, we have
\begin{eqnarray*}
a^*\tau &=& (1 + a^+)\tau \\
&=& 1\varphi + a^+\psi\\
&=& 1 + (a\psi)^+\\
&=& (a\psi)^*\\
&=&(a\tau)^*,
\end{eqnarray*}
for all $a \in H$.

%\begin{eqnarray*}
%s^*\tau &=& (x+a)^*\tau\\
%&=&  ((x^*a)x^*)\tau\\
%&=& (x^* + (x^*a)^+x^*)\tau\\
%&=& x^*\varphi + ((x^*a)^+x^*)\psi\\
%&=& (x\varphi)^* + ((x\varphi)^*(a\psi))^+(x\varphi)^*\\
%&=& (x\varphi + a\psi)^*\\
%& = &(s\tau)^*.
%\end{eqnarray*}

Since the definition of $\tau$ was forced, it is unique.
The proof is complete.
\eop

The semiring $\N$ has a natural bi-action on any hemiring $H$
with $na = an$ defined as the $n$-fold sum of $a$ with itself, 
for all $n \in \N$ and $a \in H$. 
The following fact was shown in \cite{DrosteKuich}.

\begin{prop}
\label{prop-n}
In any Conway hemiring $H$, 
\begin{eqnarray*}
(na)^+n &=& n(an)^+\\
(an)^+a &=& a(na)^+,
\end{eqnarray*}
\end{prop}
for all $n\in \N$ and $a \in H$. It follows that $(an)^*a = a(na)^*$ 
for all $n\in \N$ and $a \in H$.

\begin{thm}
\label{cor-free1}
\label{thm-free1}
The free partial Conway semiring generated by a Conway hemi\-ring 
$H$ is $(\N\os H, H, ^*)$.
\end{thm}

{\sl Proof.} Suppose that $(S',I',^*)$ is a partial Conway semiring 
and let $\psi : H \to I'$ be a Conway hemiring morphism (where
$I'$ is equipped with the plus operation determined by the 
star operation on $I$).
We want to prove that 
there is a unique partial Conway semiring morphism $\psi^\sharp : 
(\N\os H, H , ^*) \to (S',I',^*)$ extending $\psi$.

Let $\varphi$ denote the unique semiring morphism $\N \to S'$.
Then, by Proposition~\ref{prop-n},
 the assumptions of Theorem~\ref{thm-cop1} hold.  
By that theorem, there is a unique morphism $\tau
: (\N  \os H, H,^*)  \to (S',I',^*)$ of partial Conway semirings extending 
both $\varphi$ and $\psi$.
The morphism $\psi^\sharp = \tau$ is the required extension 
of $\psi$. It is clear that the extension is unique. 
\eop 

\begin{cor}
\label{cor-embed}
Every Conway hemiring embeds in a partial Conway semiring.
\end{cor}

The forgetful functor from partial Conway semirings to Conway hemirings
maps a partial Conway semiring $(S,I,^*)$ to the Conway
hemiring $I$ equipped with the plus operation determined by the star
operation. Moreover, this functor maps a morphism 
$\tau: (S,I,^*) \to (S',I',^*)$ to its restriction to $I$, 
viewed as a function $I \to I'$. A concrete representation of
the left adjoint of this functor is 
provided by Theorem~\ref{cor-free1}.
It maps a Conway hemiring $H$ to the partial Conway semiring
$(\N\os H,H,^*)$, and a hemiring morphism $\psi: H \to H'$ 
to the partial Conway semiring morphism $\tau : (\N \os H, H,^*)
\to (\N \os H', H',^*)$, which in turn maps $n + a$ to $n + a\psi$
for all $n \in \N$ and $a \in H$.  
 
Let $\B$ denote the Boolean semiring. Note that when $H$ is 
an idempotent Conway hemiring, then $\B$ has the natural 
bi-action on $H$ given by $1a = a1 = a$ and $0a = a0 = 0$,
for all $a \in H$. Clearly, (\ref{eq-action}) holds.

\begin{cor}
Suppose that $H$ is an idempotent Conway hemiring.
Then the free idempotent partial Conway semiring 
generated by $H$ is $(\B \oplus H, H, ^*)$.
\end{cor}

{\sl Proof.} The proof is the same as above, one needs to replace 
$\N$ by $\B$. \eop

\section{Extending a Conway hemiring with a Conway semiring}
\label{sec-extConway}

Suppose that $H$ is a Conway hemiring and $S_0$ is a Conway semiring
with a bi-action of $S_0$ on $H$ which 
satisfies (\ref{eq-action}). Our aim in this section is to turn $S_0\os H$ 
into a Conway semiring with a total star operation and to prove a result
analogous to Theorem~\ref{thm-cop1}. But first we need a result of independent 
interest. 

\begin{thm}
\label{thm-ext}
Suppose that $S$ is a semiring with a distinguished ideal $I$
and a subsemiring $S_0$ such that $S$ is the direct sum 
of $S_0$ and $I$, so that each $s\in S$ may be written in a unique 
way as $s = x + a$ with $x\in S_0$ and $a \in I$. 
Suppose that $S_0$ is a Conway semiring with a 
star operation $^*: S_0 \to S_0$, and that $(S,I,^*)$ is a partial 
Conway semiring with a star operation $^*: I \to S$. Then 
there is a unique way of extending both star 
operations to an operation $^*: S \to S$ such that 
$S$ becomes a Conway semiring. 
%When $S_0$ is an iteration 
%semiring and $(S,I,^*)$ is a partial iteration semiring,
%then so is $S$ equipped with the extended star operation.
\end{thm}

{\sl Proof.} 
Our proof is similar to that of the Matrix Extension Theorem in \cite{BEbook}.
Since $S$ is the direct sum of $S_0$ and $I$, $S_0 \cap I = \{0\}$. 
Since $S_0$ is a Conway semiring, we have $0^* = 1$ in $S_0$, and 
since $(S,I,^*)$ is a partial Conway semiring, $0^* = 1$ also in $(S,I,^*)$. 
Thus it is legitimate to use the same notation for both 
star operations. 

Below we will follow the subsequent notational convention: $s,s_1,s_2$ will
denote arbitrary elements of $S$, $x,y$ elements of $S_0$ and $a,b$ 
elements of $I$.  In order to define the star operation on $S$, suppose 
that $s \in S$ with $s = x + a$. We define $s^* = (x^*a)^*x^*$. 
Since $x \in S_0$ and $x^*a \in I$, $x^*$  and $(x^*a)^*$ exist
and our definition makes sense. Since $0^* = 1$, 
the new star operation extends the original ones. But we still have to 
prove that, equipped with this star operation, $S$ is a Conway 
semiring. We will often use without mention the fact that 
if $s_1$ or $s_2$ is in $I$, then $(s_1s_2)^*s_1 = s_1(s_2s_1)^*$.

First we establish the fixed point identity that will be used in the 
proof of the product star identity. Let $s = x + a$. Then
\begin{eqnarray*}
ss^* + 1 
&=& (x+a)(x+a)^* + 1\\
&=& (x+a)(x^*a)^*x^* + 1\\
&=& x(x^*a)^* x^* + a(x^*a)^*x^* + 1\\
&=& xx^*(ax^*)^* + ax^*(ax^*)^* + 1\\
&=& xx^*(ax^*)^* + (ax^*)^* \\
&=& (xx^* + 1)(ax^*)^*\\
&=& x^*(ax^*)^*\\
&=& (x^*a)^*x^*\\
&=& (x+ a)^*\\
&=& s.
\end{eqnarray*}

Next, we want to prove that $(s_1 + s_2)^* = (s_1^*s_2)^*s_1^*$ 
holds for all $s_1,s_2 \in S$. Let $s_1 = x+a$ and $s_2 = y+ b$. 

{\em Case 1.} $s_1 = a$ and $s_2 = y$. 
Then 
\begin{eqnarray*}
(s_1 + s_2)^* 
& = & (y+a)^*\\
& = & (y^*a)^*y^* \\
 & = & y^*(ay^*)^*\\
 & = & y^*(a + ay^+)^* \\
 & = & y^*(a^*ay^+)^*a^*\\
 & = & y^*(a^+yy^*)^*a^* \\
 & = & (y^*a^+y)^*y^*a^* \\
 & = & (y + a^+y)^*a^* \\
 & = & (a^*y)^*a^* \\
 & = & (s_1^*s_2)^*s_1^*. 
\end{eqnarray*}

{\em Case 2}. $s_1 = x + a$ and $s_2 = b$. Then
\begin{eqnarray*}
(s_1 + s_2)^*  
 &=& (x + a + b)^*  \\
 &=& (x^*(a+b))^*x^* \\
 &=& (x^*a + x^*b)^*x^* \\
 & = & ((x^*a)^*x^*b)^*(x^*a)^*x^*\\
 & = & ((x+a)^*b)^*(x+a)^* \\
 & = & (s_1^*s_2)^*s_1^*. 
\end{eqnarray*}  

{\em Case 3.} $s_1= x+a$ and $s_2 = y$. 
\begin{eqnarray*}
(s_1 + s_2)^* 
&= &  ((x+y) + a)^* \\
 & = & ((x+y)^*a)^*(x+y)^* \\
 & = & ((x^*y)^*x^*a)^*(x^*y)^*x^*\\
 & = & (x^*y + x^*a)^*x^* \\
 & = & (x^*a + x^*y)^*x^* \\
 & = & ((x^*a)^*x^*y)^*(x^*a)^*x^*\\
 & = & ((x+a)^*y)^*(x+a)^* \\
 & = & (s_1^*s_2)^*s_1^*,
\end{eqnarray*} 
where we used Case 1.  The last case is the general one.

{\em Case 4.} $s_1 = x+a$ and $s_2 = y + b.$
\begin{eqnarray*}
(s_1 + s_2)^* & = & ((x+a+y) + b)^* \\
 & = & ((x+a+y)^*b)^*(x+a+y)^*\\
 & = & (((x+a)^*y)^*(x+a)^*b)^*((x+a)^*y)^*(x+a)^*\\
 & = & ((x+a)^*y + (x+a)^*b)^*(x+a)^* \\
 & = & (s_1^*s_2)^*s_1^*.  
\end{eqnarray*}   
where we used Case 2 twice and Case 3 once.

Our last task is to prove the identity $(s_1s_2)^*s_1 = s_1(s_2s_1)^*$
that holds by assumption 
 when both $s_1$ and $s_2$ are in $S_0$, or one of 
$s_1$ and $s_2$ is in $I$. We have 3 cases.

{\em Case 1.} $s_1= x$ and $s_2 = y + b$. We have
\begin{eqnarray*}
(s_1s_2)^*s_1 &=& (x(y+b))^*x \\
&=& (xy + xb)^*x\\
&=& ((xy)^*xb)^*(xy)^*x\\
&=& (x(yx)^*b)^*x(yx)^*\\
&=& x((yx)^*bx)^*(yx)^*\\
&=& x(yx + bx)^*\\
&=& x((y+b)x)^*\\
&=& s_1(s_2s_1)^*.
\end{eqnarray*}

{\em Case 2.} $s_1 = x+a$, $s_2 = y$. Now, 
using the fixed point identity 
in the second line and Case 1 in the third, 
\begin{eqnarray*}
(s_1s_2)^*s_1 &=& ((x+a)y)^*(x+a)\\
&=& ((x+a)y ((x+a)y)^* + 1) (x+a)\\
&=& ((x+a)(y(x+a))^*y + 1)(x+a)\\
&=& (x+a)(y(x+a))^*y(x+a) + (x+a)\\
&=& (x+a)( (y(x + a)^*y(x+a) + 1)\\
&=& (x+a)(y(x+a))^*\\
&=& s_1(s_2s_1)^*.
\end{eqnarray*}

Last, we consider the general case, $s_1 = s$ and $s_2 = y+b$. Then
\begin{eqnarray*}
(s_1s_2)^* s_1 &=& (s(y+b))^*s \\
&=& ((sy)^*sb)^*(sy)^*s\\
&=& (s(ys)^*b)^*s(ys)^*\\
&=& s((ys)^*bs)^*(ys)^*\\
&=& s(ys + bs)^*\\
&=& s((y+b)s)^*\\
&=& s_1(s_2s_1)^*.
\end{eqnarray*} 
The proof of the theorem is complete. \eop

\begin{remark}
Theorem~\ref{thm-ext} holds (with the same proof) if we replace 
the assumption that $S$ is the direct sum of $S_0$ and $I$ 
by the assumption that 
\begin{itemize}
\item each element of $S$ can be written as a sum $x + a$ 
with $x \in S_0$ and $a \in I$, and 
\item for all $x,y\in S_0$ and $a,b\in I$,
if $x + a = y + b$, then $(x^*a)^*x^* = (y^*b)^*y^*$. 
\end{itemize}
Moreover, if the first condition holds, then the second one is also necessary.
\end{remark}

\begin{thm}
\label{thm-cop2}
Suppose that $H$ is a Conway hemiring and $S_0$ is a Conway semiring.
Moreover, suppose that there is a bi-action of $S_0$ on $H$ which 
satisfies (\ref{eq-action}). Then there is a unique way to turn 
$S_0 \oplus H$ into a Conway semiring such that the star operation 
extends the one on $S_0$ and the plus operation determined by the 
star operation extends the plus operation of $H$. 

Moreover, $S_0 \oplus H$ has the following universal property.
Suppose that $S'$ is a Conway semiring, $I'$ is an ideal of $S'$,
$\varphi: S_0 \to S'$ is a Conway semiring morphism and $\psi: 
H \to I'$ is a Conway hemiring morphism satisfying the compatibility 
conditions (\ref{eq-compatible1}) and (\ref{eq-compatible2}).
%\begin{eqnarray}
%\label{eq-compatible action1}
%(x\varphi)(a\psi) &=& (xa)\psi\\
%\label{eq-compatible action2}
%(a\psi)(x\varphi) &=& (ax)\psi
%((x\varphi)(a\psi))^+(x\varphi) &=& (x\varphi)((a\psi)(x\varphi))^+\\
%((a\psi)(x\varphi))^+(a\varphi) &=& (a\psi)((x\varphi)(a\varphi))^+
%\end{eqnarray}
%for all $x \in S_0$ and $a \in H$.
Then there is a unique Conway semiring morphism
$\tau: S_0 \oplus H \to S'$ such that $\kappa\tau = \varphi$ and 
$\lambda\tau = \psi$, i.e., such that $\tau$ extends both $\varphi$ 
and $\psi$.    
\end{thm}

{\sl Proof.}
As shown in Proposition~\ref{prop-ext}, $(S_0 \oplus H,H,^*)$, the 
extension of $H$ by $S_0$, is a partial Conway semiring with 
distinguished ideal $H$. The star operation on $H$ is the one 
determined by the original plus operation of $H$ as a Conway hemiring.
By Theorem~\ref{thm-ext}, we can further extend the star operations on $S_0$ 
and $H$ to a single star operation on $S_0 \os H$ in such a way that $S_0 \os H$
becomes a Conway semiring with the required properties. 
The uniqueness of the extension is clear. 

We already know that there is a unique 
semiring morphism $\varphi^\sharp: S_0\oplus H \to S'$ which extends 
both $\varphi$ and $\psi$, namely the function 
mapping $s= x+a$ with $x \in S_0$ and $a \in H$ to 
$x\varphi + a \psi$. Our remaining task is to show that 
$\tau$ preserves star. To prove this, let $s = x + a$
as above. We have:
\begin{eqnarray*}
s^*\tau &=& (x+a)^*\tau\\
&=&  ((x^*a)^*x^*)\tau\\
&=& (x^* + (x^*a)^+x^*)\tau\\
&=&  x^*\varphi + ((x^*a)^+x^*)\psi\\
&=& (x\varphi)^*  + ((x^*a)^+\psi)(x^*\varphi)\\
&=& (x\varphi)^* + ((x^*a)\psi)^+(x\varphi)^*\\
&=& (x\varphi)^* + ((x^*\varphi)(a\psi))^+(x\varphi)^*\\
&=& (x\varphi)^* + ((x\varphi)^*(a\psi))^+(x\varphi)^*\\
&=& (x\varphi + a\psi)^*\\
& = &(s\tau)^*. \eop 
\end{eqnarray*}

We end this section by presenting an application of Theorem~\ref{thm-cop2}. 
Note that every Conway hemiring $H$ satisfying $1^+ = 1$ is 
idempotent, since $1 + 1 = 1^+ + 1 = 1^+ = 1$ by the plus fixed 
point identity. It follows that every Conway semiring 
satisfying $1^* = 1$ is idempotent. The semiring $\B$ is naturally and uniquely equipped
with a Conway semiring (in fact, iteration semiring) structure, letting 
$0^* = 1^* = 1$. 

%First let $S_0$ 
%be the Conway semiring (in fact iteration semiring, see below) 
%$\N_\infty = \N\cup \{\infty\}$ obtained from $\N$ by adding a 
%point at infinity. In this Conway semiring, $0^* = 1$ 
%and $x^* = \infty$, for all $x \neq 0$.

\begin{cor}
Suppose that $H$ is an idempotent Conway hemiring. Then there is a unique
way to turn $\B \oplus H$ into an (idempotent) Conway semiring, by defining 
$(x + a)^* = a^*$, for all $x \in \B$ and $a \in H$. Moreover, $\B \oplus H$ 
has the following universal property. Suppose that $S'$ is an 
idempotent Conway semiring, $I'$ is an ideal of $S'$, 
$\varphi: \B \to S'$ is a Conway semiring morphism and $\psi: 
H \to I'$ is a Conway hemiring morphism satisfying the compatibility 
conditions 
(\ref{eq-compatible1}) and (\ref{eq-compatible2}).
Then there is a unique Conway semiring morphism
$\tau: \B \oplus H \to S'$ such that $\kappa\tau = \varphi$ and 
$\lambda\tau = \psi$.
\end{cor}

{\sl Proof.} This follows from Theorem~\ref{thm-cop2} by noting that 
(\ref{eq-compatible1}) and (\ref{eq-compatible2}) hold automatically.
\eop

\section{Iteration hemirings}

In order to obtain a complete description of the equational properties of the
regular languages, Conway \cite{Conway} associated an identity with each 
finite simple group.  Iteration semirings \cite{Esiteration} are Conway semirings satisfying
the group identities (or the commutative identities \cite{BEbook}). 
The completeness of the iteration semiring identities together with the 
identity $1^* = 1$ for regular languages was proved in \cite{Krob}. 
Iteration semirings, and the closely related iterative semirings
\cite{Esiteration} equipped with a partially defined star operation, 
also play a fundamental role in the axiomatization results of 
\cite{BErational,EKsemialgebras,Salomaa}. The class of iteration
semirings includes the continuous idempotent semirings and
Kozen's Kleene algebras \cite{Kozen}.

In this section, we introduce 
iteration hemirings and iterative hemirings and use them to 
provide complete axiomatizations for regular languages and rational power 
series equipped with the plus operation.  

When $S$ is a Conway semiring, so is the semiring $S^{n \times n}$ 
of all $n \times n$ matrices for each $n\geq 1$, where the star 
of a matrix is defined by the well-known \emph{matrix star formula} \cite{BEbook,Conway}:
\begin{align}
\label{matrix star id}
\begin{pmatrix}
X & Y \\
U & V
\end{pmatrix}^*
&=
\begin{pmatrix}
\alpha & \beta\\
\gamma & \delta
\end{pmatrix},
\end{align}
where
%bd below I would suggest either an alignat or a gather environment.
\begin{align*}
\begin{array}{lr}
\alpha= (X + Y V^* U)^*  & \beta = \alpha Y V^*\\
\gamma= \delta U X^* & \delta = (V + UX^*Y)^*.
\end{array}
\end{align*}
It is known, cf. \cite{BEbook}, that in Conway semirings the definition of the 
star operation on $S^{n \times n}$
does not depend on how the matrix is split into four parts. 
It follows that 
\begin{align}
\label{matrix plus id}
\begin{pmatrix}
X & Y \\
U & V
\end{pmatrix}^+
&=
\begin{pmatrix}
\alpha' & \beta'\\
\gamma' & \delta'
\end{pmatrix},
\end{align}
where
%bd below I would suggest either an alignat or a gather environment.
\begin{align*}
\begin{array}{lr}
\alpha'= (X + Y V^* U)^+  & \beta' = \beta\\
\gamma'= \gamma & \delta' = (V + UX^*Y)^+.
\end{array}
\end{align*}
($\beta$ and $\gamma$ are defined as above.)

In a similar way, when $(S,I,^*)$ is a partial Conway semiring,
then so is $S^{n \times n}$ with distinguished ideal 
$I^{n \times n}$ and star operation defined by the matrix 
star formula (\ref{matrix star id}). See \cite{BET}.
Since by Corollary~\ref{cor-embed} every 
Conway hemiring embeds in a partial Conway semiring,
a similar fact holds for matrix hemirings: if $H$ is a Conway
hemiring then so is 
each matrix hemiring $H^{n \times n}$ with plus operation
defined by (\ref{matrix plus id}).

For later use, we note that the \emph{permutation identity}
holds in all Conway semi\-rings $S$: for all $M\in S^{n \times n}$ 
and $n \times n$ permutation matrix $\pi$, 
\begin{eqnarray*}
(\pi^{-1} M \pi)^* &=& \pi^{-1} M^* \pi.
\end{eqnarray*}
(Of course, a permutation matrix $\pi$ is a $0$-$1$-matrix with a
single occurrence of $1$ in each row and column, and its inverse
$\pi^{-1}$ is its transpose.)
The same identity holds in all partial Conway semirings 
$(S,I,^*)$ when each entry of $M$ is in $I$. 
In a Conway hemiring $H$, a variant of the permutation identity 
holds:
\begin{eqnarray*}
(\pi^{-1} M \pi)^+ &=& \pi^{-1} M^+ \pi.
\end{eqnarray*}
Indeed, this identity holds in all partial Conway semirings,
but every Conway hemiring embeds in a partial Conway semiring.

Conway \cite{Conway} introduced an identity associated 
with each finite group. Suppose that $G$ is a finite 
group of order $n$. Without loss of generality we may
assume that the elements of $G$ are the integers in $[n] = \{1,\ldots,n\}$
with product denoted $i \cdot j$, or just $ij$,
for all $i,j \in [n]$. 
Consider the matrix $M_G = M_G(x_1,\ldots,x_n)$, 
\begin{align}
M_G &= 
\label{matrix MG}
\begin{pmatrix}
x_{1^{-1}1} & \ldots & x_{1^{-1}n} \\
       &  \vdots &       \\
x_{n^{-1}1} & \ldots & x_{n^{-1}n}  \\
\end{pmatrix}
\end{align}
over the variables $x_1,\ldots,x_n$ associated with the group elements.
Note that the $(i,j)$th entry of $M_G$ is the variable $x_k$ associated with 
the group element $k$ such that $ik = j$ (i.e., $(M_G)_{i,j} = x_{i^{-1} j}$). We say that the 
\emph{group identity associated with $G$} holds in a 
Conway semiring $S$ if the first row of $M_G^*$ adds up to 
$(x_1 + \ldots + x_n)^*$, for each $x_1,\ldots,x_n \in S$.

Note that each row and each column of $M_G$ is a permutation of the 
first row. It is known that in Conway semirings, the same holds 
for $M_G^*$ (see also below). Thus, the satisfaction of the 
group identity associated with $G$ in a Conway semiring 
does not depend on the order by which the group elements are enumerated,
so that we may assume that integer $1$ is the unit element of $G$.  
Moreover, if the group identity 
associated with $G$ holds, then the sum of each row or column
of $M_G^*$ is $(x_1+\ldots+ x_n)^*$.

In a similar way, we say that the group identity associated with $G$ 
holds in a partial Conway semiring $(S,I,^*)$ if it holds when each
$x_i$ belongs to $I$. Finally, we say that the \emph{plus group 
identity associated with $G$} holds in a Conway hemiring $H$ if 
the first row (or any other row or column) of $M_G^+$ adds up to 
$(x_1 + \ldots + x_n)^+$, for all $x_1,\ldots,x_n\in H$. 

Following \cite{BEbook,Esiteration}, we define a \emph{(partial) iteration semiring}
to be a (partial) Conway semiring satisfying all group identities. Similarly,
we define an \emph{iteration hemiring} to be a Conway hemiring satisfying  
all plus group identities. Morphisms of (partial) iteration semirings 
are just (partial) Conway semiring morphisms, and morphisms of iteration 
hemirings are Conway hemiring morphisms. 

\begin{prop}
A Conway semiring $S$ is an iteration semiring iff, equipped with the 
plus operation $s^+ = ss^* = s^*s$ determined by the star operation,  
it is an iteration hemiring. Moreover, a semiring $S$ equipped with 
a plus operation $^+ : S \to S$ is an iteration hemiring iff it is an iteration 
semiring with star operation $^*: S \to S$ defined by $s^* = 1 + s^+$
for all $s \in S$. 
\end{prop}

{\sl Proof.} We only prove the first statement. Suppose that $S$ 
is a Conway semiring. 

When $G$ is a group of order $n$ and $x_1,\ldots,x_n$ 
are in $S$, then $M_G^* = E_n + M_G^+$, where $E_n$ is the 
$n \times n$ unit matrix. If the sum of the first row of $M_G^+$ 
is $(x_1 + \ldots + x_n)^+$, then the sum of the first row of 
$M_G^*$ is $(x_1 + \ldots + x_n)^*$. This proves that if $S$ 
is an iteration hemiring, then it is an iteration semiring.

Suppose now that $S$ is an iteration semiring, and consider 
a matrix $M_G=M_G(x_1,\ldots,x_n)$ associated with a finite 
group of $G$ order $n$. Since $M_G^+ = M_G M_G^*$, and since 
the sum of the entries of each column of $M_G^*$ is 
$(x_1 + \ldots + x_n)^*$, the sum of the entries of the first row of
$M_G^+$ is $(x_1 +\ldots + x_n)(x_1 + \ldots + x_n)^*
= (x_1 + \ldots + x_n)^+$. \eop

Examples of iteration hemirings involve the \emph{iterative hemirings}.
An iterative hemiring is a hemiring $H$, equipped with a plus operation 
$^+: H \to H$, such that for all $a,b\in H$, the fixed point equation 
$x = ax + b$ has $a^*b$ as its \emph{unique} solution. (Here, we again write 
$a^*b$ for $a^+b + b$.) Note that every hemiring morphism between iterative 
hemirings automatically preserves the plus operation.

\begin{thm}
\label{eq-iterative}
Every iterative hemiring $H$ is an iteration hemiring.
Moreover, when $H$ is an iterative hemiring, then, equipped with the 
plus operation (\ref{matrix plus id}), so is $H^{n \times n}$ 
for each $n\geq 1$.
\end{thm} 

{\sl Proof.} Suppose that $H$ is an iterative hemiring,
and consider the semiring $\N \os H$ with distinguished 
ideal $H$. For each $a \in H$, let $a^* = 1 + a^+ $ in $\N \os H$. 

{\em Claim.} For each $a \in H$ and $s \in \N \os H$,
$a^*s$ is the unique solution in $\N \os H$ of the equation 
$x = ax + s$.

Indeed, it is clear that $a^*s$ is a solution. Now let $s = m+b$
and suppose that $n + c$ is a solution, where $m,n \in \N$ and $b,c \in H$.
Then $n + c = a(n + c) + (m + b)$, i.e., $n + c = m + ac + (na + b)$
and we have $n = m$ and $c = ac + (ma + b)$. Since the equation $x = ax + (ma + b)$
has a unique solution in $H$, it follows now that $c = a^*(ma +b)$. 
Thus, $n+c = m + a^*(ma + b) = (1 + a^*a)(m + b) = a^*s$.  

We have proved that $(\N \os H, H,^*)$ is a partial iterative semiring \cite{BET}.  
Thus, as shown in \cite{BET}, $\N \oplus H$ is a partial iteration semiring.
It follows that $H$ is an iteration hemiring.

Let $S = \N \os H$. Since $(S, H, ^*)$ is a partial iterative semiring, 
so is  $(S^{n \times n}, H^{n \times n}, ^*)$, cf. \cite{BET},   
where the star operation is defined by (\ref{matrix star id}). 
It follows that equipped with the plus operation
(\ref{matrix plus id}),  $H^{n \times n}$ is an iteration hemiring.

Actually $H$ has the following property, which implies 
all plus group identities: for all matrices $M\in H^{m \times m}$,
$N \in H^{n \times n}$ and $Q \in H^{n \times m}$, if 
$MQ = QN$ then $M^+Q = QN^+$. Indeed, if $MQ = QN$, then 
$QN^+$ is a solution of the equation $X = MX + MQ$, since 
$MQN^+ + MQ = QNN^+ + QN = Q(NN^+ + N) = QN^+$. Thus 
$M^+ Q= QN^+$. \eop 

We have shown in Proposition~\ref{prop-ext} 
that if $S_0$ is a semiring and $H$ is a Conway 
hemiring with an appropriate bi-action of $S_0$, then 
$(S_0 \oplus H,H,^*)$ is a partial Conway semiring,
where the star operation is determined by the plus operation 
of the Conway hemiring $H$. The following fact is clear:

\begin{prop}
\label{prop-sum-iteration}
When $S_0$ is semiring and $H$ is an iteration hemiring 
equipped with a bi-action of $S_0$ on $H$ satisfying
(\ref{eq-action}), the partial Conway semiring 
$(S_0 \os H, H,^*)$ clearly satisfies all plus group identities 
and thus all group identities, so that it is a partial iteration 
semiring. 
\end{prop}

From Proposition~\ref{prop-sum-iteration}
and Theorem~\ref{thm-cop1} we immediately have:

\begin{cor}
\label{cor-cop3}
Suppose that $H$ is an iteration hemiring, $S_0$ is a semiring
with a bi-action on $H$ such that 
(\ref{eq-action}) holds 
for all $x \in S_0$ 
and $a \in H$. Suppose that $(S',I',^*)$ is a partial iteration semiring, 
$\varphi: S_0 \to S'$ is a semiring morphism and $\psi: 
H \to I'$ is an iteration hemiring morphism 
satisfying (\ref{eq-compatible1}) and (\ref{eq-compatible2}),
for all $x \in S_0$ and $a \in A$.
Then there is a unique partial iteration semiring morphism
$\tau: (S_0 \oplus H, H,^*) \to (S',I',^*)$ such that $\kappa\tau = \varphi$ and 
$\lambda\tau = \psi$.  
\end{cor}

\begin{cor}
\label{cor-free partial iteration}
The free partial iteration semiring generated by an iteration hemi\-ring 
$H$ is $(\N\os H, H, ^*)$.
\end{cor}

\begin{cor}
\label{cor-embed partial iteration}
Every iteration hemiring embeds in a partial iteration semiring.
\end{cor}

\begin{cor}
Suppose that $H$ is an idempotent iteration hemiring.
Then the free idempotent partial iteration semiring 
generated by $H$ is $(\B \oplus H, H, ^*)$.
\end{cor}

We now establish Theorem~\ref{thm-ext} for iteration semirings.

\begin{thm}
\label{thm-ext2}
Suppose that $S$ is a semiring with a distinguished ideal $I$
and a subsemiring $S_0$ such that $S$ is the direct sum 
of $S_0$ and $I$. Suppose that $S_0$ is an iteration semiring with a 
star operation $^*: S_0 \to S_0$, and that $(S,I,^*)$ is a partial 
iteration semiring with a star operation $^*: I \to S$. Then 
there is a unique way of extending both star 
operations to an operation $^*: S \to S$ such that 
$S$ becomes an iteration semiring. 
\end{thm}

{\sl Proof.} We already know that there is a unique way of turning 
$S$ into a Conway semiring such that the star operation extends 
both the one defined on $S_0$ and the one on $I$. We are forced to define 
$s^* = (x^*a)^*x^*$ for all $s = x + a$ with $x \in S_0$ and $a \in I$.
By Theorem~\ref{thm-ext}, equipped with this star operation,
$S$ is a Conway semiring. So it suffices to establish the 
group identities. To this end, suppose that $G$ is a group over the set $[n]$,
where $n \geq 1$, and let $s_i = x_i + a_i \in S$ with $x_i \in S_0$ 
and $a_i\in I$, for all $i \in [n]$. Let 
\begin{eqnarray*}
M &=& M_G(s_1,\ldots,s_n)\\
X &=& M_G(x_1,\ldots,x_n)\\
A &=& M_G(a_1,\ldots,a_n),
\end{eqnarray*}
so that $M = X + A$.
Then $M^* = (X^*A)^*X^*$. We have to prove that the first row of 
$M^*$ sums up to $(s_1 + \ldots + s_n)^*$.

We note the following fact. Given an $n \times n$ matrix $N$ over $S$, 
we have $N = M_G(s_1',\ldots,s_n')$ for some $s_1',\ldots,s_n'$ in $S$ 
iff 
\begin{eqnarray*}
\rho_i^{-1} N \rho_i = N
\end{eqnarray*}
for all $i \in [n]$, where $\rho_i$ is the permutation matrix 
with $(\rho_i)_{jk} = 1 $ iff $ij = k$. 
Thus, we have $\rho_i^{-1}X \rho_i = X$ and $\rho_i^{-1} A \rho_i = A$ for all $i$.
Now by the permutation identity that holds in $S_0$,
$\rho_i^{-1}X^*\rho_i = X^*$ for all $i$, and thus 
also $\rho_i^{-1} X^*A \rho_i = \rho_i^{-1}X^*\rho_i \rho_i^{-1} A \rho_i =
X^*A$ for all $i$. Moreover, since the permutation identity also holds in 
$(S,I,^*)$, it follows that 
$\rho_i^{-1} (X^*A)^* \rho_i = (X^*A)^*$, for all $i$. 
Finally, $\rho_i^{-1} (X^*A)^*X^* \rho_i = 
\rho_i^{-1} (X^*A)^* \rho_i \rho_i^{-1} X^* \rho_i = (X^*A)^*X^*$ 
for all $i$. 

Let us introduce the following notations.
\begin{eqnarray*}
x &=& x_1 + \ldots + x_n\\
a &=& a_1 + \ldots +a_n\\
s &=& x + a.
\end{eqnarray*}
Each row or column sum of $X^*$ and $A^*$ is $x^*$ and $a^*$,
respectively, since the identity associated with 
$G$ holds in $S_0$ and $(S,I,^*)$.
Also, each row or column sum of $(X^*A)^*$
is the star of the row (or column) sum of $X^*$ multiplied with the 
constant row (or column) sum of $A$, i.e., $(x^*a)^*$. 

We conclude that each row (or column) sum of $M^* = (X^*A)^*X^*$ 
is $(x^*a)^* x^* = (x+a)^* = s^*$. \eop 

\begin{remark}
Theorem~\ref{thm-ext2} remains valid if instead of the assumption 
that $S$ is the direct sum of $S_0$ and $H$ we require that each
element of $S$ can be written as a sum $x + a$ with $x \in S_0$ 
and $a \in A$, and if $x + a = y + b$ with $x,y \in S_0$ and $a,b \in A$,
then $(x^*a)^*x^* = (y^*b)^*y^*$.
\end{remark} 

Using Theorem~\ref{thm-ext2}, we have: 

\begin{cor}
\label{thm-cop22}
Suppose that $H$ is a iteration hemiring and $S_0$ is an iteration semiring.
Moreover, suppose that there is a bi-action of $S_0$ on $H$ which 
satisfies (\ref{eq-action}). Then there is a unique way to turn 
$S_0 \oplus H$ into an iteration semiring such that the star operation 
extends the one on $S_0$ and the plus operation determined by the 
star operation extends the plus operation of $H$. 

Moreover, $S_0 \oplus H$ has the following universal property.
Suppose that $S'$ is an iteration semiring, $I'$ is an ideal of $S'$,
$\varphi: S_0 \to S'$ is an iteration semiring morphism and $\psi: 
H \to I'$ is an iteration hemiring morphism satisfying the compatibility 
conditions (\ref{eq-compatible1}) and (\ref{eq-compatible2}).
%\begin{eqnarray}
%\label{eq-compatible action1}
%(x\varphi)(a\psi) &=& (xa)\psi\\
%\label{eq-compatible action2}
%(a\psi)(x\varphi) &=& (ax)\psi
%((x\varphi)(a\psi))^+(x\varphi) &=& (x\varphi)((a\psi)(x\varphi))^+\\
%((a\psi)(x\varphi))^+(a\varphi) &=& (a\psi)((x\varphi)(a\varphi))^+
%\end{eqnarray}
%for all $x \in S_0$ and $a \in H$.
Then there is a unique iteration semiring morphism
$\tau: S_0 \oplus H \to S'$ such that $\kappa\tau = \varphi$ and 
$\lambda\tau = \psi$, where $\kappa$ and $\lambda$ denote the 
natural embeddings of $S_0$ and $H$ in $S_0 \oplus H$.   
\end{cor}

\section{Free iteration hemirings}
\label{sec-free}

In this section, we combine results from the previous sections 
with some results from \cite{BErational,EKsemialgebras,Krob,Salomaa}
to provide a concrete description of the free iteration hemiring
and that of the free idempotent iteration hemiring, 
freely generated by an alphabet $A$. The description uses 
rational power series and regular languages.

Recall that when $S$ is a semiring and $A$ is an alphabet, then 
a \emph{power series} \cite{Bersteletal,KuichSalomaa} in 
$S\llangle A^* \rrangle$ is a function $f: A^* \to S$,
usually written as a formal sum $\sum_{w\in A^*}(f,w)w$.
We say that a series $f$ is \emph{proper} if $(f,\epsilon) = 0$,
where $\epsilon$ denotes the empty word. 
As usual, we equip $S\llangle A^* \rrangle$ with the 
sum and product operations defined by 
\begin{eqnarray*}
(f + g,w) &=& (f,w) + (g,w)\\
(fg,w) &=& \sum_{uv = w}(f,u)(g,v)
\end{eqnarray*}
Each semiring element $s\in S$ may be identified with the 
series that maps $\epsilon$ to $s$ and all other words to $0$.
It is well-known that, equipped with the above operations
and constants $0,1$, $S \llangle A^* \rrangle$ is a semiring.
The set of proper series is an ideal. We may identify this ideal 
with the hemiring $S \llangle A^+ \rrangle$ of all functions 
$A^+\to S$. In fact, $S\llangle A^+ \rrangle$ is an iterative hemiring, 
and hence an iteration hemiring, since fixed point equations 
over $S\llangle A^+ \rrangle$ have unique solutions 
\cite{Bersteletal,KuichSalomaa}. The plus operation 
is defined by 
\begin{eqnarray*}
(f^+,w) &=& \sum_{w = u_1\ldots u_n,\ u_i \neq \epsilon}\prod_{i = 1}^n (f,u_i).
\end{eqnarray*}
Also, $(S\llangle A^*\rrangle, S\llangle A^+\rrangle,^*)$ is a partial 
iterative and hence partial iteration semiring, where the star operation 
defined on $S\llangle A^+\rrangle$ is that determined by the above plus 
operation.

We also equip $S\llangle A^* \rrangle$ (and $S\llangle A^+ \rrangle$)
with the natural bi-action of $S$:
\begin{eqnarray*}
(sf,w) &=& s(f,w)\\
(fs,w) &=& (f,w)s
\end{eqnarray*}
for all series $f \in S\llangle A^* \rrangle$ and $s \in S$. Indeed, with the identification given above, $S$ is a subsemiring of $S \llangle A^* \rrangle$, and the bi-action of $S$ coincides with the product operation of $S \llangle A^* \rrangle$ restricted to $S$ and $S \llangle A^* \rrangle$.

\begin{lem}
We have $(sf)^+s = s(fs)^+$ for all $s \in S$ and $f \in S\llangle A^+\rrangle$.
\end{lem}

Thus, (\ref{eq-action}) holds.

Each letter $a \in A$ may be identified with the series that 
maps $a$ to $1$ and all other words to $0$.  
A series in $S\llangle A^* \rrangle$ is called {\em rational},
cf. \cite{Bersteletal,KuichSalomaa}, if it can be generated from
the series corresponding to the letters of $A$ by the semiring operations,
the natural (bi-)action of $S$, and the partially defined star operation.
We let $S^\rat\llangle A^* \rrangle$ denote the semiring 
of rational series, which, equipped with the ideal $S^\rat\llangle A^+ \rrangle$ 
of proper rational series and the star operation, is 
a partial iterative semiring and a partial iteration semiring.
Also, $S^\rat\llangle A^+ \rrangle$
is an iterative hemiring and hence an iteration hemiring, 
which is generated from the series corresponding 
to the letters in $A$ by the hemiring operations, the 
(bi-)action of $S$, and the plus operation.  

It is well-known that $\B\llangle A^* \rrangle$ is isomorphic to
the semiring of all languages in $A^*$, equipped with set union 
as sum and concatenation as product. The constants $0$ and $1$
are the empty language and the language $\{\epsilon\}$.
Also, $\B^\rat\llangle A^* \rrangle$  ($(\B^\rat\llangle A^+ \rrangle)$, resp.)
is isomorphic the the semiring of regular languages (regular languages not containing
the empty word) in $\Sigma^*$. The star and plus operations are the usual ones.

In \cite{BErational}, it was shown that for every alphabet $A$, 
the semiring $\N^\rat\llangle A^* \rrangle$, equipped with
the ideal $\N\llangle A^+\rrangle$ of proper rational series, 
and the star operation on proper series, is the free partial
iteration semiring, freely generated by $A$. 
Using this result and Theorem~\ref{thm-cop1}, we prove:

\begin{thm}
\label{thm-free iteration hemiring}
For each alphabet $A$, $\N^\rat\llangle A^+ \rrangle$ is the free 
iteration hemiring, freely generated by $A$.
\end{thm}

{\sl Proof.} We have proved that the forgetful functor from partial 
Conway semirings to Conway hemirings has as left adjoint the 
functor $\N \oplus -$ mapping a Conway hemiring to the 
partial Conway semiring $(\N \oplus H, H,^*)$, 
where the star operation is determined by the plus operation of $H$
(cf. Theorem~\ref{cor-free1}).
 When $H$ is an iteration hemiring, 
$(\N \os H,H,^*)$ is a partial iteration semiring. Thus the same 
functor, restricted to iteration hemirings, is a left adjoint 
of the forgetful functor from partial iteration semirings 
to iteration hemirings.

Suppose that $H$ is an iteration hemiring. We can
form the partial iteration 
semiring $(\N \os H,H,^*) $, as well as $(\N \os \N^\rat \llangle A^+ \rrangle,
\N^\rat\llangle A^+ \rrangle,^*)$, which is just the partial 
iteration semiring $(\N^\rat\llangle A^* \rrangle, \N^\rat\llangle A^+ \rrangle, ^*)$,
where we have identified  any series in $\N^\rat\llangle A^+ \rrangle$
with the corresponding proper series in $\N^\rat\llangle A^* \rrangle$.
Now it is known from \cite{BErational} that $(\N^\rat\llangle A^*\rrangle,
N^\rat\llangle A^+ \rrangle, ^*)$ 
is the free partial iteration semiring, freely generated by $A$. 
So given a function $\psi: A \to H$, there is a unique 
partial iteration semiring morphism 
$(\N^\rat\llangle A^* \rrangle, \N^\rat\llangle A^+ \rrangle,^*) \to 
(\N \os H, H, ^*)$ extending $\psi$. 
The restriction of this partial iteration semiring
morphism to $\N^\rat\llangle A^+\rrangle$ is the required
iteration hemiring morphism 
$\psi^\sharp : \N^\rat\llangle A^+\rrangle \to H$
extending $\psi$.  

The uniqueness of the extension follows from Corollary~\ref{cor-free1}.
\eop

\begin{cor}
For each alphabet $A$, $\N^\rat \llangle A^+ \rrangle$ is the free 
iterative hemiring, freely generated by $A$.
\end{cor}

{\sl Proof.} 
This follows from Theorem~\ref{thm-free iteration hemiring} 
since $\N^\rat\llangle A^+ \rrangle$ is an iterative hemiring
and every iterative hemiring is an iteration hemiring. \eop 

\begin{cor}
The variety of hemirings with a plus operation 
generated by the iterative hemirings is the 
class of iteration hemirings. An identity holds in all
iterative hemirings iff it holds in all iteration hemirings
iff it holds in all iteration hemirings $\N^\rat\llangle A^+\rrangle$,
or in $\N^\rat\llangle A^+ \rrangle$, where $A$ is the $2$-element alphabet. 
\end{cor}

In the idempotent case, we have:

\begin{thm}
\label{thm-free idempotent iteration hemiring}
For each alphabet $A$, $\B^\rat\llangle A^+\rrangle$ is both the 
free idempotent iteration hemiring 
and the free idempotent iterative hemiring,
 freely generated by $A$.
\end{thm} 

{\sl Proof.} The proof uses results from \cite{EKsemialgebras,Krob}
and is similar to that of Theorem~\ref{thm-free iteration hemiring}.
 \eop

\begin{cor}
The variety of hemirings with a plus operation 
generated by the idempotent iterative hemirings is the 
class of idempotent iteration hemirings.
An identity holds in all idempotent
iterative hemirings iff it holds in all idempotent iteration hemirings
iff it holds in all idempotent iteration hemirings $\B^\rat\llangle A^+\rrangle$,
or in $\B^\rat\llangle A^+ \rrangle$, where $A$ is the $2$-element alphabet. 
\end{cor}

%It has been shown in \cite{EsikKuichALGUNIV} that if $S$ is an `atomistic' and `proper' commutative %semiring, 
%which is additively generated by its multiplicative invertible elements, 
%such as $\B$ and $\N$, then $(S^\rat\llangle A^* \rrangle, S^\rat\llangle A^+\rrangle,^*)$
%is the free partial iteration semiring, freely generated by $A$. 
%A common generalization of Theorems~\ref{thm-free iteration hemiring} 
%and \ref{thm-free idempotent iteration hemiring} can be derived 
%by using this result:%

%\begin{thm}
% Let $S$ be an atomistic and proper commutative semiring, 
%which is additively generated by its multiplicative invertible elements.
%Then for each alphabet $A$, $S^\rat\llangle A^+ \rrangle$ has the following 
%universal property. Given any iteration hemiring $S'$
%\end{thm}   

\newpage

\vspace*{.2in}

\begin{center}
{\bf \Large Part 2}
\end{center}

\vspace*{.2in}

\section{Introduction to Part 2 }

In \cite{BEbook}, Conway and iteration semiring-semimodule pairs were used 
as an abstract framework for B\"uchi-automata on $\omega$-words. These structures,
consisting of a semiring $S$ acting on a semimodule $V$, are equipped with 
both a star operation $^*: S \to S$ and an omega operation $^\omega : S \to V$.
They were subsequently studied in \cite{EsikKuichsem1,EsikKuichsem2}, where 
refinements of the Kleene theorem of \cite{BEbook} were obtained. Since in
several situations, the star and omega operations cannot be made into total 
operations, partial Conway and iteration semiring-semimodule pairs were 
introduced in \cite{Esikpartial}. In this paper, we follow another line 
to deal with the problem of partiality. Instead of semiring-semimodule pairs,
we will consider hemiring-hemimodule pairs $(H,V)$, equipped with total 
operations $^+: H  \to H$ and $^\omega : H \to V$. We define Conway 
and iteration hemiring-hemimodule pairs and study their relation to 
(partial) Conway and iteration semiring-semimodule pairs. We show how to 
add freely a semiring or a Conway semiring $S_0$ to a Conway or iteration 
hemiring-hemimodule pair $(H,V,^*,^\omega)$ to obtain a partial Conway 
or iteration semiring-semimodule pair (Theorems~\ref{thm-cop3}, \ref{thm-cop5}
and Corollaries~\ref{cor-ext} and \ref{cor-cop8}),  define automata 
in Conway hemiring-hemimodule pairs, and prove a general Kleene theorem 
(Theorem~\ref{thm-Kleene}). 
In the final sections, we apply this general result to the analysis of the 
infinitary behavior of weighted transition systems (automata).  Such infinitary quantitative behaviors, computed e.g. by discounting or by average of weights, were recently introduced and investigated in \cite{Chatterjeeetal, Chatterjeeetal2} and subsequently in \cite{DrosteMeineckeMFCS, DrosteMeinecke, Meinecke}.

\section{Conway hemiring-hemimodule pairs}

In this section, we define Conway-hemiring hemimodule pairs 
as a generalization of the Conway semiring-semimodule pairs of 
\cite{BEbook}, which form an abstract framework for studying 
the infinitary behavior of finite automata, see 
\cite{BEbook,EsikKuichsem1,EsikKuichsem2}. 

Suppose that $H$ is a hemiring. A (left) \emph{$H$-hemimodule}
is a commutative monoid $V = (V,+,0)$ together with a left action
$H \times V \to V$ subject to the expected laws. A morphism of left 
$H$-semimodules is a monoid morphism that respects the action.

When $S = (S,+,\cdot,0.1)$ is a semiring, it is also a hemiring,
so that we can speak of $S$-hemimodules $V$. However,
the action may not be unitary. When it is, we call $V$ an 
\emph{$S$-semimodule}.

A \emph{Conway semiring-semimodule pair} \cite{BEbook} $(S,V,^*,^\omega)$ 
consists of a Conway semiring $S$ (equipped with a star
operation $^*: S\to S$), an $S$-semimodule $V$, and an 
\emph{omega operation} $^\omega: S \to V$, such that
the following \emph{sum omega} and \emph{product omega} 
identities hold:
\begin{eqnarray}
\label{eq-sum omega}
(x+y)^\omega &=& (x^*y)^*x^\omega + (x^*y)^\omega\\
\label{eq-product omega}
(xy)^\omega &=& x(yx)^\omega.
\end{eqnarray}
Note that the {\em omega fixed point identity}
\begin{eqnarray}
\label{eq-omega fixed point}
xx^\omega &=& x^\omega
\end{eqnarray}
is a consequence of the product omega identity,
and the {\em zero omega identity}
 \begin{eqnarray*}
 0^\omega &=& 0
 \end{eqnarray*} 
is an instance of this identity. 
In a similar fashion, a \emph{Conway hemiring-hemimodule} pair
$(H,V,^+,^\omega)$ consists of a Conway hemiring $H$, an $H$-hemimodule $V$
and an omega operation $^\omega: H \to V$ such that 
the sum omega (\ref{eq-sum omega}), product omega (\ref{eq-product omega})
and omega fixed point (\ref{eq-omega fixed point}) identities 
hold. (Recall that $x^*y$ stands 
for $x^+ y  + y$ and $yx^*$ for $y + yx^+$.
Similarly, we write $x^*v$ for $x^+v + v$, so that  
$(x^*y)^*x^\omega$ is  $(x^*y)^+x^\omega + x^\omega$,
etc.) It follows that the zero omega identity holds.
Morphisms of Conway semiring-semimodule pairs 
are given by a Conway semiring morphism and a semimodule 
morphism that jointly preserve the action and the omega operation. 
Morphisms of Conway hemiring-hemimodule pairs are defined in a similar 
fashion.

Following \cite{BET}, we also define \emph{partial 
Conway semiring-semimodule pairs} $(S,I,V,^*,^\omega)$.
These structures consist of a partial Conway semiring $(S,I,^*)$ and 
an omega operation $^\omega: I \to V$ satisfying the sum omega identity (\ref{eq-sum omega}) for 
all $x,y \in I$, and the product omega identity (\ref{eq-product omega}) when $x$ 
or $y$ is in $I$. Morphisms of partial Conway semiring-semimodule 
pairs are defined as those of Conway semiring-semimodule pairs and 
additionally preserve the distinguished ideal.

We will make use of the following fact without mention.

\begin{lem}
A semiring-semimodule pair $(S,V)$ equipped with a star operation 
$^*: S \to S$ and omega 
operation $^\omega: S\to V$
is a Conway semiring-semimodule pair iff it is a 
Conway hemiring-hemimodule pair equipped with 
the plus operation $a \mapsto a^+ = aa^* = a^*a$ and the omega operation.
\end{lem}

We will also make use of the following fact.

\begin{lem}
\label{lem-prodomega}
In a Conway hemiring-hemimodule pair $(H,V,^+,^\omega)$, we 
have
\begin{eqnarray}
\label{eq-lem1}
(x^*y)^\omega &=& x^*(yx^*)^\omega\\
\label{eq-lem2}
(yx^*)^\omega &=& y(x^*y)^\omega
\end{eqnarray}
for all $x,y \in H$.
\end{lem}

{\sl Proof.} By definition, we have 
\begin{eqnarray*}
(x^*y)^\omega &=& (x^+y + y)^\omega \\
&=& (y^*x^+y)^*y^\omega + (y^*x^+y)^\omega
\end{eqnarray*}
and 
\begin{eqnarray*}
x^*(yx^*)^\omega 
&=& x^*(y + yx^+)^\omega\\
&=& x^*(y^+x^+)^*y^\omega + x^*(y^+x^+)^\omega\\
&=& x^+(y^+x^+)^*y^\omega + (y^+x^+)^*y^\omega + x^+(y^+x^+)^\omega + (y^+x^+)^\omega.
\end{eqnarray*}
So (\ref{eq-lem1}) holds if we can show that
\begin{eqnarray*}
(y^*x^+y)^*y^\omega &=& x^+(y^+x^+)^*y^\omega + (y^+x^+)^*y^\omega\\
(y^*x^+y)^\omega &=& x^+(y^+x^+)^\omega + (y^+x^+)^\omega.
\end{eqnarray*}
But 
\begin{eqnarray*}
(y^*x^+y)^*y^\omega &=&
y^*x^+(y^+x^+)^*yy^\omega + y^\omega \\
&=& (y^+x^+)^+y^\omega + x^+(y^+x^+)^*y^\omega + y^\omega\\
&=& (y^+x^+)^*y^\omega + x^+(y^+x^+)^*y^\omega
\end{eqnarray*}
and
\begin{eqnarray*}
(y^*x^+y)^\omega &=& y^*x^+(y^+x^+)^\omega\\
&=& y^+x^+(y^+x^+)^\omega  + x^+(y^+x^+)^\omega\\
&=& (y^+x^+)^\omega  + x^+(y^+x^+)^\omega.
\end{eqnarray*}
The proof of (\ref{eq-lem1}) is complete. 
As for (\ref{eq-lem2}), we use the omega fixed point identity (\ref{eq-omega fixed point}) 
and (\ref{eq-lem1}):
\begin{eqnarray*}
(yx^*)^\omega &=& yx^*(yx^*)^\omega\\
&=&y(x^*y)^\omega.
\eop 
\end{eqnarray*}

\section{Extending a Conway hemiring-hemimodule pair with a 
semi\-ring}

In Section~\ref{sec-extend}, we showed how to add freely a semiring to a 
Conway hemiring to obtain a partial Conway semiring. 
Here, our aim is to add a semiring to a Conway hemiring-hemimodule pair.

Let $(H,V,^+,^\omega)$ be a Conway hemiring-hemimodule pair, and suppose that 
$S_0$ is a semiring with a bi-action on $H$ satisfying (\ref{eq-action}) and 
 one of 
\begin{eqnarray}
\label{eq-action2}
(xa)^\omega = x(ax)^\omega\quad {\rm and}\quad (ax)^\omega = a(xa)^\omega
\end{eqnarray} 
for all $x \in S_0$ and $a \in H$. (We note that if the first identity holds, 
then so does the second using the omega fixed point identity, and vice versa.)

In Section~\ref{sec-extend}, we have constructed the partial Conway semiring $(S_0 \os H, H , ^*)$
and showed that it has a certain universal property (cf. Theorem~\ref{thm-cop1}).
Suppose now that $S_0$ also has a unitary left action on $V$, i.e., 
$V$ is also an  $S_0$-semimodule, and that this action is compatible 
with the left action of $S_0$ on $H$:
\begin{eqnarray}
\label{eq-I}
(xa)v &=& x(av)\\
\label{eq-II}
(ax)v &=& a(xv),
\end{eqnarray}
for all $x \in S_0$, $a \in H$ and $v \in V$.
Then $S_0 \os H$ has the natural unitary left action on $V$, defined by:
\begin{eqnarray*}
sv &=& xv + av,
\end{eqnarray*}
for all $s = x + a$ in $S_0 \os H$ with $x \in S_0$ and $a \in H$,
and for all $v$ in $V$. 
Here we only check that $s(s'v) = (ss')v$ for all $s,s' \in S_0 \os H$
and $v \in V$. Let $s = x + a$ as before, and similarly, let $s' = y + b$.
Then:
\begin{eqnarray*}
s(s'v) &=& s(yv + bv) \\
&=& x(yv + bv) + a(yv + bv)\\
&=& x(yv) + x(bv) + a(yv) + a(bv)\\
&=& (xy)v + ((xb)v + (ay)v + (ab)v)\\
&=& (xy)v + (xb + ay + ab)v\\
&=& (ss')v.
\end{eqnarray*}

\begin{prop}
\label{prop-sum-Conway-pair}
Under the above assumptions, 
$(S_0\os H, H, V, ^*,^\omega)$ is a partial Conway semiring-semimodule pair. 
\end{prop}

{\sl Proof.} The only difficulty is to show that the product omega 
identity holds. To this end, let $s = y + b \in S_0 \os H$ 
with $y \in S_0$ and $b \in H$, and let $a \in H$. 
Then:
\begin{eqnarray*}
(as)^\omega &=& (a(y + b))^\omega \\
&=& (ay + ab)^\omega \\
&=& ((ay)^*ab)^*(ay)^\omega + ((ay)^*ab)^\omega\\
&=& (a(ya)^*b)^*a(ya)^\omega + (a(ya)^*b)^\omega\\
&=& a((ya)^*ba)^*(ya)^\omega + a((ya)^*ba)^\omega\\
&=& a((ya)^*ba)^*(ya)^\omega + ((ya)^*ba)^\omega)\\
&=& a(ya + ba)^\omega\\
&=& a((y+b)a)^\omega \\
&=& a(sa)^\omega.
\end{eqnarray*}

Also,
\begin{eqnarray*}
(sa)^\omega &=& ((y+b)a)^\omega\\
&=& (ya + ba)^\omega\\
&=& (ya + ba)(ya + ba)^\omega\\
&=& ya(ya + ba)^\omega + ba(ya + ba)^\omega
\end{eqnarray*}
and 
\begin{eqnarray*}
s(as)^\omega &=& (b+ y)(a(b + y))^\omega \\
&=& y (ay + ab)^\omega + b(ay + ab)^\omega. 
\end{eqnarray*}
So to conclude that the omega product identity holds,
we have to verify that
\begin{eqnarray*}
(ay + ab)^\omega &=& a(ya + ba)^\omega.
\end{eqnarray*}
However,
\begin{eqnarray*}
a(ya + ba)^\omega 
&=& 
a((ya)^*ba)^*(ya)^\omega + a((ya)^*ba)^\omega\\
&=& 
(a(ya)^*b))^*a(ya)^\omega + (a(ya)^*b)^\omega\\
&=& 
((ay)^*ab)^*(ay)^\omega + ((ay)^*ab)^\omega\\
&=& 
(ay + ab)^\omega. \eop 
\end{eqnarray*}

\begin{thm}
\label{thm-cop3}
Suppose that $(H,V,^+,^\omega)$ is a Conway hemiring-hemimodule
pair, $S_0$ is a semiring
with a bi-action on $H$ and a left action on $V$ such that 
(\ref{eq-action}), (\ref{eq-action2}), (\ref{eq-I}) and (\ref{eq-II})  hold  
for all $x \in S_0$, $a \in H$ and $v \in V$.
Then the partial Conway semiring-semimodule pair
$(S_0\os H, H, V, ^*,^\omega)$ has the following universal property.
Suppose that $(S',I',V',^*,^\omega)$ 
is a partial Conway semiring-semimodule pair,  
$\varphi: S_0 \to S'$ is a semiring morphism, and $\psi = (\psi_H,\psi_V)$
is a Conway hemiring-hemimodule morphism 
$(H,V,^+,^\omega) \to (I',V',^+,^\omega)$ (where $^+$ 
is the plus operation determined by the star operation
of $(S',I',V',^*,^\omega)$)
with $\psi_H: H \to I'$ 
and $\psi_V: V \to V'$ such that 
\begin{eqnarray}
\label{eq-mor1}
(x\varphi)(a\psi_H) &=& (xa)\psi_H\\
\label{eq-mor2}
(a\psi_H)(x\varphi) &=& (ax)\psi_H\\
\label{eq-mor3}
(x\varphi)v &=& (xv)\psi_V,
\end{eqnarray}
for all $x \in S_0$, $a \in A$ and $v \in V$, so that $(\varphi,\psi_H)$ 
is a morphism $(S_0,V) \to (S',V')$ of semiring-semimodule pairs. 
Then there is a unique partial Conway semiring morphism
$\tau =(\tau_S,\tau_V): (S_0 \oplus H,H,V,^*,^\omega)  \to (S',I',V',^*,^\omega)$ 
extending $\varphi$ and $\psi$. 
\end{thm}

{\sl Proof.} We already know that $(S_0 \os H,H,V,^*,^\omega)$ 
is a partial Conway semiring-semimodule pair. 
Given $\varphi$, $\psi_H$ and $\psi_V$, 
we define $s\tau_S = x\varphi + a\psi$ 
for all $s = x + a \in S_0 \os H$ with $x \in S_0$ 
and $a \in H$, and we define $v\tau_V = v\psi_V$ 
for all $v \in V$, i.e., $\tau_V = \psi_V$.
We need to show that $\tau$ is a morphism of
partial Conway semiring-semimodule pairs extending $\varphi$ and $\psi$.

It is clear that $\tau_S$ extends $\varphi_H$ and $\psi_H$
and we already know that $\tau_S$ preserves the semiring operations
and constants, the distinguished ideal, and star. 
Since $\tau_V = \psi_V$, to complete the proof,
we only need to verify that 
$\tau_S$ and $\tau_V$ preserve the action.
Let $s = x + a \in S_0 \oplus H$ with $x \in S_0$ and $a \in H$.

{\em Claim}: $\tau_S$ and $\tau_V$ preserve the action. 
Below we will just write $\tau$ for both $\tau_S$ and $\tau_V$,
and similarly for $\psi$. 
\begin{eqnarray*}
(sv)\tau &=& ((x+a)v)\tau\\
&=& (xv + av)\tau\\
&=& (xv)\tau + (av)\tau\\
&=& (x\varphi)(v\psi) +(a\psi)(v\psi)\\
&=& (x\varphi +a\psi)(v\psi)\\
&=& ((x+a)\tau)(v\psi)\\
&=& (s\tau)(v\psi).
\end{eqnarray*}

Since the definition of $\tau$ was forced, it is unique.
The proof is complete.
\eop

When $(H,V,^+,^\omega)$ is a Conway hemiring-hemimodule pair,
$\N$ has a natural bi-action on $H$ and a natural left action 
on $V$. It was shown in \cite{DrosteKuich} that the action
on $H$ satisfies (\ref{eq-action}). It is clear
that (\ref{eq-I}) and (\ref{eq-II}) hold. We show that 
(\ref{eq-action2}) holds.

\begin{lem}
Suppose that $(H,V,^*,^\omega)$ is a Conway
hemiring-hemimodule pair. Then for any 
$n\in \N$ and $a\in H$,
\begin{eqnarray*}
(an)^\omega = a(na)^\omega \quad
{\rm and}
\quad 
(na)^\omega = n(an)^\omega.
\end{eqnarray*}
\end{lem}

{\sl Proof.} 
We prove the first identity by induction on $n$.
When $n = 0$, both sides are $0$. In the induction
step, suppose that $n = m + 1$ and that our claim holds for $m$.
Then, using Proposition~\ref{prop-n} to go from the second line to the third
and the induction hypothesis to go from the third line to the fourth,   
\begin{eqnarray*}
a(na)^\omega &=& a(ma + a)^\omega \\
&=& a((ma)^*a)^\omega + a((ma)^*a)^*(ma)^\omega\\
&=& (a(ma)^*)^\omega + ((ma)^*a)^*a(ma)^\omega\\
&=& ((am)^*a)^\omega + ((am)^*a)^*(am)^\omega\\
&=& (am + a)^\omega \\
&=& (na)^\omega. 
\end{eqnarray*}
To complete the proof, we show that the first identity implies the second,
as long as the omega fixed point identity holds. Indeed, if these hold, then 
\begin{eqnarray*}
(na)^\omega &=& na(na)^\omega\\
&=& n(an)^\omega. \eop 
\end{eqnarray*}

\begin{cor}
Suppose that $(H,V,^+,^\omega)$ is a Conway hemiring-hemimodule pair.
Then $(\N \oplus H,H,^*)$ is partial Conway semiring-semimodule pair,
where the star operation on $H$ is determined by the plus operation
of $H$. Moreover, $(\N \os H, H, V, ^*, ^\omega)$ has the following universal property.
Given any partial Conway semiring-semimodule pair
$(S',I',V',^*,^\omega)$, any morphism of Conway hemiring-hemimodule
pairs $\psi = (\psi_H,\psi_V): (H,V,^+,^\omega) \to 
(I',V',^+,^\omega)$ (where $(I',V',^+,^\omega)$ is equipped with the
plus operation determined by the star operation $^*: I' \to S'$), there is a
unique morphism $\tau = (\tau_S,\tau_V): 
(\N \os H, H,V,^*,^\omega) \to (S',I',V',^*,^\omega)$ 
extending $\psi$.
\end{cor}

{\sl Proof.} This follows from Theorem \ref{thm-cop3} using the previous lemma 
and noting that (\ref{eq-mor1}),
(\ref{eq-mor2}) and (\ref{eq-mor3}) hold. \eop  

\begin{cor}
\label{emb-Conway-module}
Every Conway hemiring-hemimodule pair embeds in a partial 
Conway semiring-semi\-module pair. 
\end{cor}

\begin{cor}
Suppose that $(H,V,^+,^\omega)$ is an idempotent Conway hemiring-hemimodule pair.
Then $(\B \oplus H,H,^*)$ is a partial Conway semiring-semimodule pair 
which has the following universal property.
Given any partial idempotent Conway semiring-semimodule pair
$(S',I',V',^*,^\omega)$, any morphism of Conway hemiring-hemimodule
pairs $\psi = (\psi_H,\psi_V): (H,V,^+,^\omega) \to (I',V',^+,^\omega)$ 
(where $(I',V',^+,^\omega)$ is equipped with the
plus operation determined by the star operation $^*: I \to S'$), there is a
unique morphism $(\B \os H, H,V,^*,^\omega) \to (S',I',V',^*,^\omega)$ 
extending $(\psi_H,\psi_V)$.
\end{cor}

\section{Extending a Conway hemiring-hemimodule with a 
Conway semiring}

In this section, we extend the results of Section~\ref{sec-extConway} 
to Conway hemiring-hemimodule pairs. We will make use of the following 
improvement of the Matricial Extension Theorem of \cite{BEbook}.

\begin{thm}
Suppose that $(S,I,V,^*,^\omega)$ is a partial Conway 
semiring-semimodule pair and $(S_0,V,^*,^\omega)$ is a 
Conway semiring-semimodule pair.
Suppose that $S$ is the direct sum of $S_0$ and $I$.
Then there is a unique way of extending the star and omega 
operations to $S$ such that $(S,V,^*,^\omega)$ becomes 
a Conway semiring-semimodule pair.
\end{thm}

{\sl Proof.}
First let us notice that $S_0 \cap I = \{0\}$, and that 
$0^\omega =  0$ for both omega operations.

We have already proved that, under the assumptions, there is a unique 
way of extending the star operation to $S$ such that it becomes a
Conway semiring. One needs to define $(x+a)^*  = (x^*a)^*x^*$
for all $x \in S_0$ and $a \in A$. Consider now the omega operation.
When $s = x + a$ with $x \in S_0$ and $a \in I$, we are forced to
define $s^\omega = (x^*a)^*x^\omega + (x^*a)^\omega$. It is clear that 
this definition extends both the omega operation defined on $S_0$
and the one defined on $I$. But we still have to show that the 
sum omega and product omega identities hold. 

We use the following notational convention: $s,s_1,s_2$ denote elements 
of $S$, $x,y$ are in $S_0$ and $a,b$ in $I$.

First we establish the omega fixed point identity. Let $s = x + a$. 
\begin{eqnarray*}
ss^\omega & = & (x + a)((x^*a)^*x^\omega + (x^*a)^\omega) \\
 & = & (x+a)(x^*a)^*x^\omega + (x+a)(x^*a)^\omega.
\end{eqnarray*}
We show
\begin{eqnarray*}
(x+a)(x^*a)^*x^\omega & = & (x^*a)^*x^\omega
\end{eqnarray*}and
\begin{eqnarray*}
(x+a)(x^*a)^\omega & = & (x^*a)^\omega .
\end{eqnarray*}
Indeed,
\begin{eqnarray*}
x(x^*a)^*x^\omega + a(x^*a)^*x^\omega  & = & x( (x^*a)(x^*a)^* + 1)x^\omega 
+ a(x^*a)^*x^\omega \\
 & = & (x^*a(x^*a)^* + 1)x^\omega \\
 & = & (x^*a)^*x^\omega.
\end{eqnarray*}
Also
\begin{eqnarray*}
x(x^*a)^\omega + a(x^*a)^\omega & = & xx^*(ax^*)^\omega + (ax^*)^\omega 
\\
 & = & x^*(ax^*)^\omega \\
 & = & (x^*a)^\omega.
\end{eqnarray*}
Thus,
\begin{eqnarray*}
(x+a)(x^*a)^*x^\omega + (x+a)(x^*a)^\omega & = & (x^*a)^*x^\omega 
+ (x^*a)^\omega \\
 & = & (x+a)^\omega.
\end{eqnarray*}

Next we prove that the sum omega identity holds.
We will make use of the following identities that hold in 
all Conway semirings: 
\begin{eqnarray}
(s_1^*s_2)^* & = & s_2^*(s_1^+s_2^+)^* \label{suba}\\
s_2^*(s_1^+s_2^+)^* & = &  s_1^*s_2^+(s_1^+s_2^+)^* + 1 \label{subb}\\
(s_1^*s_2)^* & = & s_1^*s_2^+(s_1^+s_2^+)^* + 1.\label{subc}
\end{eqnarray} 
Let $s_1 =  x + a$ and $s_2 = y + b$.

{\sc Case 1.} $s_1 = a$ and $s_2 = y.$
\begin{eqnarray*}
(s_1 + s_2)^\omega  & = & (a + y)^\omega  \\
& = & (y + a)^\omega \\
 & = & (y^*a)^*y^\omega  + (y^*a)^\omega \\
 & = & (y^*a)^*y^\omega  + (a + y^+a)^\omega \\
 & = & (y^*a)^*y^\omega + (a^*y^+a)^*a^\omega + (a^*y^+a)^\omega.
\end{eqnarray*} 
Also, 
\begin{eqnarray*}
(s_1^*s_2)^*s_1^\omega  + (s_1^*s_2)^\omega  & = & (a^*y)^*a^\omega + (a^*y)^\omega 
 \\
 & = & (a^+y + y)^*a^\omega  + (a^+y + y)^\omega  \\
 & = & (y^*a^+y)^*y^*a^\omega + (y^*a^+y)^*y^\omega  + (y^*a^+y)^\omega 
.
\end{eqnarray*}
The proof of this case will be completed once we prove the following 
identities.
\begin{eqnarray}
\label{eqa}
(y^*a^+y)^*y^*a^\omega  & = & (a^*y^+a)^*a^\omega \\
\label{eqb}
(y^*a^+y)^*y^\omega  & = & (y^*a)^*y^\omega \\
\label{eqc}
(y^*a^+y)^\omega  & = & (a^*y^+a)^\omega .
\end{eqnarray}
{\sl Proof of} (\ref{eqa}).
\begin{eqnarray*}
(y^*a^+y)^*y^*a^\omega  & = & (1 + a^*y^+(a^+y^+)^*)a^\omega 
\end{eqnarray*}by (\ref{subb}),
\begin{eqnarray*}
 & = & a^\omega  + a^*y^+(a^+y^+)^*a^\omega \\
 & = & a^\omega  + a^*y^+(a^+y^+)^*a a^\omega \\
 & = & (1 + a^*y^+(a^+y^+)^*a)a^\omega \\
 & = & (1 + (a^*y^+a)(a^*y^+a)^*)a^\omega \\
 & = & (a^*y^+a)^*a^\omega .
\end{eqnarray*}
{\sl Proof of} (\ref{eqb}).
\begin{eqnarray*}
(y^*a^+y)^*y^\omega  & = & y^*a^+(yy^*a^+)^*yy^\omega + y^\omega \\
 & = & (y^*a^+(y^+ a^+)^* + 1)y^\omega \\
 & = & (y^*a)^*y^\omega ,
\end{eqnarray*}
by (\ref{subc}) above.\\
{\sl Proof of} (\ref{eqc}).  
\begin{eqnarray*}
(y^*a^+y)^\omega  & = & y^*(a^+y^+)^\omega  \\
 & = & y^+(a^+y^+)^\omega  + (a^+y^+)^\omega \\
 & = & (y^+a^+)^\omega  + a^+(y^+a^+)^\omega \\
 & = & a^*(y^+a^+)^\omega \\
 & = & (a^*y^+a)^\omega . 
\end{eqnarray*}
The proof of this case is complete.

%%%%%%%%%%%%%%%%%%%%%%%%%%%%%%%%%%%%%%%
{\sc Case 2.} $s_1 = x + a$ and $s_2 = b$.
\begin{eqnarray*}
\lefteqn{
(s_1^*s_2)^*s_1^\omega + (s_1^*s_2)^\omega}\\  
& = & ((x+a)^*b)^*(x+a)^\omega  
+ ((x+a)^*b)^\omega  \\
 & = & ((x^*a)^*x^*b)^*[(x^*a)^*x^\omega  + (x^*a)^\omega ] + ((x^*a)^*x^*b)^\omega 
\\
 & = & ((x^*a)^*x^*b)^*(x^*a)^*x^\omega  + ((x^*a)^*x^*b)^*(x^*a)^\omega 
 + ((x^*a)^*x^*b)^\omega \\
 & = & (x^*a + x^*b)^*x^\omega + (x^*a + x^*b)^\omega \\
 & = & (x^*(a+b))^*x^\omega + (x^*(a+b))^\omega \\
 & = & (x + (a+b))^\omega \\
 & = & ((x + a) + b)^\omega \\
 & = & (s_1 + s_2)^\omega .
\end{eqnarray*}

{\sc Case 3.} $s_1 = x$ and $s_2 = y + b$.
\begin{eqnarray*}
\lefteqn{
(s_1+s_2)^\omega }\\ 
 & = & (x + (y+b))^\omega \\
 & = & ((x+y) + b)^\omega \\
 & = & ((x+y)^*b)^*(x+y)^\omega +((x+y)^*b)^\omega  \\
 & = & ((x^*y)^*x^*b)^*((x^*y)^*x^\omega + (x^*y)^\omega ) +
          ((x^*y)^*x^*b)^\omega \\
 & = & ((x^*y)^*x^*b)^*(x^*y)^*x^\omega + ((x^*y)^*x^*b)^*(x^*y)^\omega 
+ ((x^*y)^*x^*b)^\omega \\
 & = & (x^*y + x^*b)^*x^\omega + (x^*y + x^*b)^\omega \\
 & = & (s_1^*s_2)^*s_1^\omega + (s_1^*s_2)^\omega .
\end{eqnarray*}

{\sc Case 4.} $s_1 = x + a$ and $s_2 = y$.  In the third line we use Case 3, and we use 
Case 1 in the fifth line.
\begin{eqnarray*}
\lefteqn{
(s_1 + s_2)^\omega}\\ 
 & = & ((x+a) + y)^\omega \\
 & = & (x + (a+y))^\omega \\
 & = & (x^*(a+y))^*x^\omega  + (x^*(a+y))^\omega \\
 & = & (x^*a + x^*y)^*x^\omega + (x^*a + x^*y)^\omega \\
 & = & ((x^*a)^*x^*y)^*(x^*a)^*x^\omega +
      ((x^*a)^*x^*y)^*(x^*a)^\omega + ((x^*a)^*x^*y)^\omega \\
 & = & ((x^*a)^*x^*y)^*((x^*a)^*x^\omega + (x^*a)^\omega ) +
      ((x^*a)^*x^*y)^\omega \\
 & = & ((x + a)^*y)^*(x + a)^\omega + ((x+a)^*y)^\omega \\
 & = & (s_1^*s_2)^*s_1^\omega + (s_1^*s_2)^\omega .
\end{eqnarray*}

{\sc Case 5.}  The general case.
\begin{eqnarray*}
(s_1 + s_2)^\omega  & = & ((s_1 + y) + b)^\omega \\
 & = & ((s_1+y)^*b)^*(s_1 + y)^\omega  + ((s_1 + y)^*b)^\omega 
\end{eqnarray*}
by Case 2,
\begin{eqnarray*}
 & = & ((s_1^*y)^*s_1^*b)^*(s_1+y)^\omega  + ((s_1^*y)^*s_1^*b)^\omega  \\
 & = & ((s_1^*y)^*s_1^*b)^*((s_1^*y)^*s_1^\omega + (s_1^*y)^\omega ) +  
		((s_1^*y)^*s_1^*b)^\omega 
\end{eqnarray*}
by Case 4,
\begin{eqnarray*}
 & = & ((s_1^*y)^*s_1^*b)^*(s_1^*y)^*s_1^\omega  + ((s_1^*y)^*s_1^*b)^*(s_1^*y)^\omega 
+ ((s_1^*y)^*s_1^*b)^\omega \\
 & = & (s_1^*y + s_1^*b)^*s_1^\omega  + (s_1^*y +s_1^* b)^\omega ,
\end{eqnarray*}
by Case 2,
\begin{eqnarray*}
 = (s_1^*s_2)^*s_1^\omega + (s_1^*s_2)^\omega .
\end{eqnarray*}
The proof of the sum omega identity is complete.
%%%%%%%%%%%%%%%%%%%%%%%%%%%%%%%%%%%%%%%%%%%%%%%

Last, we establish the product star identity.
Let $s_1 = x + a$, $s_2 = y + b$.
There are only three cases to consider.

{\sc Case 1.} $s_1 = x$ and $s_2 = y + b$.
\begin{eqnarray*}
(s_1s_2)^\omega & = & (x(y + b))^\omega \\
 & = & (xy + xb)^\omega \\
 & = & ((xy)^*xb)^*(xy)^\omega + ((xy)^*xb)^\omega \\
 & = & (x(yx)^*b)^*(xy)^\omega + (x(yx)^*b)^\omega \\
 & = & (x(yx)^*b)^*x(yx)^\omega + x((yx)^*bx)^\omega \\
 & = & x((yx)^*bx)^*(yx)^\omega + x ((yx)^*bx)^\omega \\
 & = & x(yx + bx)^\omega \\
 & = & s_2(s_1s_2)^\omega.
\end{eqnarray*} 

{\sc Case 2.} $s_1 = x + a$ and $s_2 = y$.  In the first line we use 
the omega fixed point identity and in the second line we use Case 1.   
\begin{eqnarray*}
((x + a)y)^\omega & = & (x+a)y((x+a)y)^\omega \\
 & = & (x+a)(y(x+a))^\omega \\
 & = & s_1(s_2s_1)^\omega .
\end{eqnarray*}

{\sc Case 3.} The general case.
\begin{eqnarray*}
(s_1s_2)^\omega  & = & (s_1y + s_1b)^\omega \\
 & = & ((s_1y)^*s_1b)^*(s_1y)^\omega + ((s_1y)^*s_1b)^\omega\\
 & = & (s_1(ys_1)^*b)^*s_1(ys_1)^\omega  + s_1((ys_1)^*bs_1)^\omega ,
\end{eqnarray*}
by Case 2,
\begin{eqnarray*}
 & = & s_1((ys_1)^*bs_1)^*(ys_1)^\omega + s_1((ys_1)^*bs_1)^\omega \\
 & = & s_1[((ys_1)^*bs_1)^*(ys_1)^\omega + ((ys_1)^*bs_1)^\omega ] \\
 & = & s_1(ys_1 + bs_1)^\omega\ =\ s_1(s_2s_1)^\omega.
\end{eqnarray*}
The proof of the Theorem is complete.
\eop 

\begin{remark}
The same result holds under the assumption that each element of $S$ 
is a sum $x + a$ with $x\in S_0$ and $a \in A$, and for
all $x,y \in S_0$ and $a,b\in A$, if $x + a = y + b$, then
$(x^*a)^*x^* = (y^*b)^*y^*$ and $(x^*a)^*x^\omega + (x^*a)^\omega 
= (y^*b)^*y^\omega + (y^*b)^\omega$. If $S$ is the direct sum
of $S_0$ and $H$, then these conditions clearly hold. 
\end{remark}

Let $(H,V,^+,^\omega)$ be a Conway hemiring-hemimodule pair
and let $S_0$ be a Conway semiring.
Moreover, suppose that there is a bi-action of $S_0$ on $H$ which 
satisfies (\ref{eq-action}) and (\ref{eq-action2}), as well as a 
left action of $S_0$ on $V$ which satisfies (\ref{eq-I}) and 
(\ref{eq-II}).  As shown above, the 
extension of $(H,V,^*,^\omega)$ by $S_0$, $(S_0 \oplus H,H,
V,^*,^\omega)$ 
is a partial Conway semiring-semimodule pair with distinguished ideal $H$.
By Theorem~\ref{thm-ext2}, we may extend the star operation on $S_0$ 
and the star operation determined by the
plus operation on $H$ to a single star operation on $S_0\os H$ 
so that it becomes a Conway semiring. Moreover, we may extend the 
omega operation defined on $S_0$ and on $H$ to $S_0\os H$ 
such that  $(S_0 \oplus H,V,^*,^\omega)$ becomes a 
Conway semiring-semimodule pair.
Recall that $S_0$ and $H$ embed in $S_0 \os H$ by $\kappa$ 
and $\lambda$.

\begin{thm}
\label{thm-cop5}
Suppose that $(H,V,^+,^\omega)$ is a Conway hemiring-hemimodule 
pair and $S_0$ is a Conway semiring.
Suppose that there is a bi-action of $S_0$ on $H$ which 
satisfies (\ref{eq-action}) and (\ref{eq-action2}), as well as a 
left action of $S_0$ on $V$ which satisfies (\ref{eq-I}) and 
(\ref{eq-II}), so that we have the 
Conway semiring-semimodule pair $(S_0 \oplus H,V,^*,^\omega)$.
 Then there is a unique way to extend the star and omega 
operation to $S_0 \os H$ so that $(S_0 \oplus H,V,^*,^\omega)$
becomes a Conway semiring-semimodule pair.

Suppose that $(S',V',^*,^\omega)$ is a Conway semiring-semimodule 
pair, $\varphi: S_0 \to S'$ is a Conway semiring morphism,  
$\psi = (\psi_S,\psi_V)$ is a Conway hemiring-hemimodule 
morphism morphism such that (\ref{eq-mor1}),
(\ref{eq-mor2}) and (\ref{eq-mor3}) hold.  
%so that   
%\begin{eqnarray*}
%(x\varphi)(a\psi_H) &=& (xa)\psi_H
%((x\varphi)(a\psi))^+(x\varphi) &=& (x\varphi)((a\psi)(x\varphi))^+\\
%((a\psi)(x\varphi))^+(a\varphi) &=& (a\psi)((x\varphi)(a\varphi))^+
%\end{eqnarray*}
%for all $x \in S_0$ and $a \in A$.
Then there is a unique Conway semiring-semimodule morphism 
$\tau = (\tau_S,\tau_V): (S_0 \oplus H,V,^*,^\omega)
\to (S',V',^*,^\omega)$ extending $\varphi$ and $\psi$. 
\end{thm}

{\sl Proof.} 
We already know (cf. Theorem~\ref{thm-cop3}) that there is a unique 
partial Conway semiring-semimodule 
morphism $\tau = (\tau_S,\tau_V):  (S_0\oplus H, H, V,^*,^\omega) \to 
(S',S',V',^*,^\omega)$ which extends 
$\varphi$ and $\psi$. In fact, $\tau_V = \psi_V$ 
and $s\tau_S = x\varphi + a\psi_H$ 
for all $s = x+a$ with $x \in S_0$ and $a \in H$.
Moreover, $\tau_S$ preserves the star operation 
on $S_0 \os H$. 

Our task is to show that the omega operation
is preserved.  To prove this, let $s = x + a$
as above. Then writing just $\tau$ for both $\tau_S$ and $\tau_V$
and just $\psi$ for $\psi_H$ and $\psi_V$, we have
\begin{eqnarray*}
s^\omega\tau &=& (x+a)^\omega\tau\\
&=&  ((x^*a)^*x^\omega + (x^*a)^\omega)\tau\\
&=&  ((x\varphi)^* a\psi)^* (x\varphi)^\omega
      + ((x\varphi)^* a \psi)^\omega\\
&=& (x\varphi + a\psi)^\omega\\
&=& (s\tau)^\omega. \eop 
\end{eqnarray*}

%%%%%%%%%%%%%%%%%%%%%%%%%%%%%%%%%% 

\section{Iteration hemiring-hemimodule pairs}

Suppose that $(S,V)$ is a semiring-semimodule pair.
Then for each $n \geq 1$, we may define the action 
of $S^{n \times n}$ on $V^n$ by $(Mv)_i = 
\sum_{j = 1}^n M_{i,j}v_j$. Equipped with this action, 
$(S^{n \times n}, V^n)$ is also a semiring-semimodule pair.

When $(S,V,^*,^\omega)$ is a Conway semiring-semimodule pair, 
we have already turned $S^{n \times n}$ into a Conway semiring. 
Following \cite{BEbook}, we may define an omega operation 
$^\omega :  S^{n \times n} \to V^n$ and obtain a 
Conway semiring-semimodule pair  $(S^{n \times n},V^n,^*,^\omega)$
for each $n \geq 1$. 
For each 

\begin{align}
\label{matrix omega id}
M &= 
\begin{pmatrix}
X & Y \\
U & V
\end{pmatrix},
\end{align}
we define 
\begin{align}
M^ \omega &=
\begin{pmatrix}
\alpha \\
\beta
\end{pmatrix},
\end{align}
where
%bd below I would suggest either an alignat or a gather environment.
\begin{align*}
\begin{array}{lr}
\alpha= (X + Y V^* U)^*YV^\omega +  (X + Y V^* U)^\omega\\
\beta = (V + UX^*Y)^*UX^\omega + (V + UX^*Y)^\omega
\end{array}
\end{align*}
It is known, cf. \cite{BEbook}, that the definition of the 
omega operation 
does not depend on how the $n \times n$ matrix $M$ is split into four parts.

In a similar way, when $(S,I,V,^*,^\omega)$ is a partial Conway semiring-semimodule pair,
then so is $$(S^{n \times n}, I^{n \times n}, V^n, ^*,^\omega)$$ with distinguished ideal 
$I^{n \times n}$ and star and omega operations defined by the matrix 
star formula (\ref{matrix star id}) and the \emph{matrix omega formula}
(\ref{matrix omega id}).
Since by Corollary~\ref{emb-Conway-module}, 
every Conway hemiring-hemimodule pair embeds in a partial Conway semiring-semimodule 
pair, a similar fact holds for matrix hemiring-hemimodule pairs: 
if $(H,V,^+,^\omega)$ is a Conway
hemiring-hemimodule pair, then so is 
$(H^{n \times n}, V^n,^+,^\omega)$ with plus operation
defined by (\ref{matrix plus id}) and omega operation defined by (\ref{matrix omega id}).  
The action of $H^{n \times n}$ on $V^n$ is defined as above.

It is known that the \emph{omega permutation identity} 
holds in all Conway semi\-ring-semimodule pairs $(S,V,^*,^\omega)$: 
for all $M\in S^{n \times n}$ and $n \times n$ permutation matrix $\pi$, 
\begin{eqnarray*}
(\pi^{-1} M \pi)^\omega &=& \pi^{-1} M^\omega.
\end{eqnarray*}
The same identity holds in all partial Conway semiring-semimodule pairs 
$(S,I,V,^*,^\omega)$ when each entry of $M$ is in $I$, and in all 
 Conway hemiring-hemimodule pairs $(H,V,^+,^\omega)$,
since every Conway hemiring-hemimodule pair embeds in a 
partial Conway semiring-semimodule pair.

Suppose that $G$ is a finite group of order $n$. 
Recall the definition of the matrix $M_G = M_G(x_1,\ldots,x_n)$.
We say that the \emph{omega group identity associated with $G$} holds in a 
Conway semiring-semimodule pair $(S,V,^*,^\omega)$, or more generally, in a 
Conway hemiring-hemimodule pair $(H,V,^*,^\omega)$, if the first 
entry of  $M_G^\omega$ is 
$(x_1 + \ldots + x_n)^\omega$, for each $x_1,\ldots,x_n$ in $S$, or in $H$.
It follows then that each entry of $M_G^\omega$ is $(x_1+ \cdots + x_n)^\omega$.

In a similar way, we say that the omega group identity associated with $G$ 
holds in a partial Conway semiring-semimodule pair $(S,I,V,^*,^\omega)$ 
if it holds when each $x_i$ belongs to $I$.

Following \cite{BEbook,Esiteration}, we define a \emph{(partial) iteration semiring-semimodule pair}
to be a (partial) Conway semiring-semimodule pair satisfying all group identities
for the star and omega operations. Similarly,
we define an \emph{iteration hemiring-hemimodule pair} to be a Conway hemiring-hemimodule 
pair satisfying  
all group identities for the plus and omega operations. Morphism of (partial) iteration 
semiring-semimodule pairs 
are just (partial) Conway semiring-semimodule pair morphisms, and morphisms of iteration 
hemiring-hemimodule pairs are Conway hemiring-hemimodule pair morphisms. 
From Lemma~\ref{lem-first} we clearly have:

\begin{lem}
\label{lem-second}
A Conway semiring-semimodule pair is an iteration semiring-semimodule pair iff 
it is an iteration hemiring-hemimodule pair. 
\end{lem}

Suppose now that $H$ is a hemiring and $V$ is a nontrivial $H$-hemimodule,
so that $V \neq \{0\}$. Moreover, suppose that $V$ is positive, so that if $v + v' = 0$
or $av = 0$, for some $v,v' \in V$ and $a\in H$ with $a \neq 0$, then $v = 0$. Then $H$ is also positive, 
since if $a + b = 0$, for some $a,b\in H$, and if $v \neq 0$,  then by $av + bv = (a+b)v =0$
we have $av = 0$ and $a = 0$. We say that $(H,V)$ is an iterative hemiring-hemimodule pair 
if $H$ is an iterative hemiring and for all $a \in H$ and $v \in V$, the equation $x = ax + v$ 
has either $x = 0$ as its unique solution, when $a = 0$ and $v = 0$, or it has a unique 
nonzero solution. A morphism of iterative hemiring-hemimodule pairs 
is just a hemiring-hemimodule pair morphism.

Suppose that $(H,V)$ is an iterative hemiring-hemimodule pair. Then as shown above,
$H$ is an iteration hemiring with plus operation such that  $a^+ = aa^+ + a$
for all $a \in H$. We also define an omega operation. Let $a \in H$. If $a = 0$, then 
$a^\omega = 0$. Otherwise, $a^\omega$ is the unique \emph{nonzero} solution in $V$ 
of the equation $x = ax$.

\begin{prop}
Every iterative hemiring-hemimodule pair is an iteration hemiring-hemimodule pair.
Every morphism of iterative hemiring-hemimodule pairs is an iteration hemiring-hemimodule 
pair morphism. 
\end{prop}

{\sl Proof.} In \cite{Esikpartial}, an iterative semiring-semimodule pair is defined
as a system $(S,I,V,^*,^\omega)$ consisting of a semiring $S$ with an ideal $I \subseteq S$, 
a positive $S$-semimodule $V$ and a star and an omega operation such that for all 
$a \in I$, $b \in S$ and $v \in V$, the equation $x = ax + b$ has $a^*b$ 
as its unique solution, and the equation $x = ax + v$ has either $0$ as its unique solution when
$a= 0$ and $v = 0$, or its unique solution is $a^\omega + a^*v$. (Thus, when $a \neq 0$, $a^\omega$ 
is the unique nonzero solution of $x = ax$ in $V$.) In particular, $(S,I,^*)$ is a partial 
iterative semiring. Also, as shown in \cite{Esikpartial}, every partial iterative 
semiring-semimodule pair is a partial iteration semiring-semimodule pair.  Thus, if $(H,V,^+,^\omega)$ 
is an iterative hemiring-hemimodule pair, then $(\N \oplus H, H, V,^*,^\omega)$ is a 
partial iterative semiring-semimodule pair and a partial iteration semiring-semimodule pair,
where $^*$ is the star operation determined by the plus operation of $(H,V,^+,^\omega)$. 
It follows that $(H,V,^+,^\omega)$ is an iteration hemiring-hemimodule pair, proving the 
first statement. The second statement is obvious. \eop

Suppose that $S_0$ is semiring and $(H,V, ^+,^\omega)$ is an iteration
hemiring-hemimodule pair. Moreover, suppose that $S_0$ has 
an appropriate bi-action on $H$ and a left action on $V$. 
Then the partial Conway semiring-semimodule pair  
$(S_0 \os H, H,V,^*,^\omega)$ is clearly  
a partial iteration semiring-semimodule pair. 
By Theorem~\ref{thm-cop3} and Lemma~\ref{lem-second}, we have:

\begin{cor}
\label{cor-ext}
Suppose that $(H,V,^+,^\omega)$ is an iteration hemiring-hemimodule 
pair, $S_0$ is a semiring with a bi-action on $H$ such that 
(\ref{eq-action}), (\ref{eq-action2}), (\ref{eq-I}) and 
(\ref{eq-II})  hold for all $x \in S_0$, 
$a \in H$ and $v \in V$. Then $(S_0 \oplus H, H, V,^*,^\omega)$ 
is partial iteration hemiring-hemimodule pair.
Suppose that $(S',I',V',^*,^\omega)$ is a partial 
iteration semiring-semimodule pair,
$\varphi: S_0 \to S'$ is a semiring morphism and $\psi = (\psi_H,\psi_V)$
is an iteration hemiring-hemimodule pair  morphism 
$(H,V,^+,^\omega) \to (I',V',^+,^\omega)$, where $^+$ is the plus operation
determined by the star operation on $I'$. Moreover, suppose that 
(\ref{eq-mor1}), (\ref{eq-mor2}) and (\ref{eq-mor3}) hold 
for all $x \in S_0$, $a \in A$ and $v \in V$.
Then there is a unique partial iteration semiring-semimodule morphism
$\tau = (\tau_S,\tau_V): (S_0 \oplus H, H,V, ^*,^\omega) \to (S',I',V',^*,^\omega)$ 
extending $\varphi$ and $\psi$.
\end{cor}

In the rest of this section, our aim is to prove that when 
$(H,V,^+,^\omega)$ is an iteration hemiring-hemimodule pair and $S_0$ is an 
iteration semiring with an appropriate bi-action on $H$ and left action on $V$,
then the Conway semiring-semimodule pair $(S_0 \os H, V,^*,^\omega)$ of 
Theorem~\ref{thm-cop5} is an iteration hemiring-hemimodule pair. But first we prove:

\begin{thm}
\label{thm-ext4}
Suppose that $(S,I,V,^*,^\omega)$ is a partial iteration semiring-semimodule
pair and $S_0$ is a subsemiring of $S$, and moreover, $(S_0,V.^*,^\omega)$ 
is an iteration semiring-semimodule pair (so that $S_0$ is an iteration semiring).
Suppose that $S$ is the direct sum of $S_0$ and $I$. Then there is a unique way 
of extending both star operations to an operation $^*: S \to S$ and the omega operations
to an operation $^\omega: S \to V$ such that 
$(S,V,^*,^\omega)$ becomes an iteration semiring-semimodule pair. 
\end{thm}

{\sl Proof.} We know that there is a unique way of turning 
$(S,V,^*,^\omega)$  into a Conway semiring-semimodule 
pair  such that the star and omega operations extend
the ones defined on $S_0$ and $I$. We are forced to define 
$s^* = (x^*a)^*x^*$  and $s^\omega = (x^*a)^*x^\omega + (x^*a)^\omega$
for all $s = x + a$ with $x \in S_0$ and $a \in I$.
From Theorem~\ref{thm-ext2} we also know that $S$, equipped 
with this star operation, is an iteration semiring.
To complete the proof, we need to show how to extend the omega operations.

To this end, suppose that $G$ is a group over the set $[n]$,
where $n \geq 1$, and let $s_i = x_i + a_i \in S$ with $x_i \in S_0$ 
and $a_i\in I$, for all $i \in [n]$. Let 
\begin{eqnarray*}
M &=& M_G(s_1,\ldots,s_n)\\
X &=& M_G(x_1,\ldots,x_n)\\
A &=& M_G(a_1,\ldots,a_n),
\end{eqnarray*}
so that $M = X + A$ and $M^\omega = (X^*A)^*X^\omega + (X^*A)^\omega$.

Let us introduce the following notations.
\begin{eqnarray*}
x &=& x_1 + \ldots + x_n\\
a &=& a_1 + \ldots +a_n\\
s &=& x + a.
\end{eqnarray*}
We already know that each row or column sum of $X^*$ and $X^*A$ is $x^*$ and $x^*a$,
respectively, since the identity associated with 
$G$ holds in $S_0$ and $(S,I,^*)$.
Each row or column sum of $(X^*A)^*$
is the star of the constant row (or column) sum of $X^*$ multiplied with the constant row
(or column) sum of $A$, i.e., $(x^*a)^*$. Also, each entry of 
$X^\omega$ is $x^\omega$, and each entry of $(X^*A)^\omega$ 
is $(x^*a)^\omega$. 

We conclude that each entry of $M^\omega = (X^*A)^*X^\omega + (X^*A)^\omega$ 
is $(x^*a)^* x^\omega + (x^*a)^\omega  = (x+a)^\omega = s^\omega$. \eop 

By Theorem~\ref{thm-ext4} and Theorem~\ref{thm-cop5}, we conclude:

\begin{cor}
\label{cor-cop8}
Suppose that $(H,V,^+,^\omega)$ is an iteration hemiring-hemimodule 
pair and $S_0$ is an iteration semiring.
Suppose that there is a bi-action of $S_0$ on $H$ which 
satisfies (\ref{eq-action}) and (\ref{eq-action2}), as well as a 
left action of $S_0$ on $V$ which satisfies (\ref{eq-I}) and 
(\ref{eq-II}). Then there is a unique way to extend the star and omega 
operation to $S_0 \os H$  so that $(S_0 \oplus H,V,^*,^\omega)$
becomes an iteration semiring-semimodule pair.

Suppose that $(S',V',^*,^\omega)$ is a iteration semiring-semimodule 
pair, $\varphi: S_0 \to S'$ is an iteration semiring morphism,  
$\psi = (\psi_S,\psi_V)$ is an iteration hemiring-hemimodule 
morphism morphism such that (\ref{eq-mor1}),
(\ref{eq-mor2}) and (\ref{eq-mor3}) hold.  
%so that   
%\begin{eqnarray*}
%(x\varphi)(a\psi_H) &=& (xa)\psi_H
%((x\varphi)(a\psi))^+(x\varphi) &=& (x\varphi)((a\psi)(x\varphi))^+\\
%((a\psi)(x\varphi))^+(a\varphi) &=& (a\psi)((x\varphi)(a\varphi))^+
%\end{eqnarray*}
%for all $x \in S_0$ and $a \in A$.
Then there is a unique iteration semiring-semimodule morphism 
$\tau = (\tau_S,\tau_V): (S_0 \oplus H,V,^*,^\omega)
\to (S',V',^*,^\omega)$ extending $\varphi$ and $\psi$. 
\end{cor}

\section{Automata}
\label{sec-automata}

Automata in Conway semirings and Conway semiring-semimodule 
pairs were defined in \cite{BEbook}. This general notion of automata 
was later refined in \cite{EsikKuichsem1,EsikKuichsem2}
and extended to partial Conway semirings and partial 
Conway semiring-semimodule pairs in \cite{BET,Esikpartial}.
We now define automata in Conway hemiring-hemimodule pairs
and relate them to automata in (partial) Conway semiring-semimodule pairs. 

Suppose that $(H,V,^+,^\omega)$ is a Conway hemiring-hemimodule 
pair and $S_0$ is a semiring with a bi-action on $H$ and a left 
action on $V$, subject to the compatibility conditions (\ref{eq-action}),
(\ref{eq-action2}), (\ref{eq-I}), (\ref{eq-II}) of the 
previous sections. Suppose that $A\subseteq H$. 
An \emph{$S_0$-automaton over $A$} in $(H,V,^+,^\omega)$ is a system $\cA = (\alpha, M, \beta,k)$,
where $\alpha \in S_0^{1 \times n}$ is the \emph{initial vector}, 
$M \in S_0\langle A \rangle ^{n \times n}$ is the \emph{transition
matrix}, $\beta\in S_0^{n \times 1}$ is the \emph{final vector},
and $k \leq n$ is an integer. Here, $S_0\langle A \rangle$ is the collection 
of all those elements of $S_0 \os H$ which are finite linear combinations 
of elements of $A$ with coefficients in  $S_0$. Note that 
$S_0\langle A \rangle \subseteq H$.
We call $n$ the \emph{dimension} of $\cA$.

The \emph{finitary behavior} of $\cA$ in $(H,V,^+,^\omega)$ 
is $|\cA|_f = \alpha M^+ \beta$, 
where we use the bi-action of $S_0$ on $H$. 
The \emph{infinitary behavior}
of $\cA$ in $(H,V,^+,^\omega)$ is $|\cA|_\omega = \alpha M^{\omega,k}$, where 
we use the left action of $S_0$ on $V$ and 
$M^{\omega,k}$ is defined in the following way.
First, split $M$ into blocks 
\begin{align}
\begin{pmatrix}
X & Y \\
U & V
\end{pmatrix},
\end{align}
where $X$ is $k \times k$, etc, and then 
\begin{align}
M^{\omega,k}&=
\begin{pmatrix}
(X + YV^*U)^\omega \\
V^*U(X + YV^*U)^\omega 
\end{pmatrix}.
\end{align}

Call an element of $H$ \emph{$S_0$-rational over $A$} in $(H,V,^+,^\omega)$, or just \emph{rational},
if it can be generated from the elements of $A$ by the rational operations of sum, 
product, plus and the bi-action of $S_0$. (See also Section~\ref{sec-free}.) 
Moreover, call an element of $V$ \emph{$S_0$-rational over $A$} in $(H,V,^+,^\omega)$,
or just \emph{rational}, 
if it can be generated 
from the elements of $A$ by the above operations, the left action 
of $S_0$ and $H$ on $V$, and omega power. 

\begin{expl}
\label{expl-buchi}
Suppose that $A$ is an alphabet. Then the hemiring $\B\llangle A^+ \rrangle$ 
may be conveniently identified with the hemiring $P(A^+)$ of all subsets of 
$A^+$, equipped with set union as sum and concatenation as product. 
Let $A^\omega$ denote the set of all $\omega$-words (sequences) over $A$,
and let $P(A^\omega)$ be the set of all subsets of $A^\omega$. Equipped  
with the operation of set union as sum and the empty set as $0$, and the 
action $LU = \{uv : u \in L,\ v \in U\}$ for all $L \subseteq A^+$ and 
$U \subseteq A^\omega$, we have a hemiring-hemimodule pair $(P(A^+),P(A^\omega))$.
The plus and omega operations are given as usual:
\begin{eqnarray*}
L^+ &=& \{u_1\ldots u_n : n > 0,\ u_1,\ldots,u_n \in L\}\\
L^\omega &=& \{u_1u_2\ldots : u_i \in L\}.
\end{eqnarray*}
(See also Section~\ref{sec-free}.) Now $(P(A^+),(P^\omega),^+,^\omega)$ 
is a Conway hemiring-hemimodule pair (since it embeds in the Conway semiring-semimodule 
pair $(P(A^*),P(A^\omega), ^*,^\omega)$ defined similarly, cf. \cite{BEbook}). 
Let $S_0 = \B$ act on $P(A^+)$ and $P(A^\omega)$ in the expected way.
Then all assumptions hold, so that if we regard $A$ as a subset of $P(A^+)$,
then we can define automata $\cA = (\alpha, M, \beta,k)$ over $A$. 
When the dimension of $\cA$ is $n$, then $\cA$ corresponds to the 
ordinary nondeterministic finite automaton (nfa) with state set $Q = \{q_1,\ldots,q_n\}$, say,
and transition function $q_j \in \delta(q_i,a)$ iff $a \in M_{ij}$, for all 
$i,j \in [n]$ and $a \in A$. A state $q_i$ is initial iff $\alpha_i = 1$, 
and final iff $\beta_i = 1$. The finitary behavior is the language 
of nonempty words accepted by the nfa. The infinitary behavior 
consists of those $\omega$-words over $A$ which have a run starting
in an initial state that visits the set $\{q_1,\ldots,q_k\}$ infinitely often,
i.e., those accepted by the B\"uchi-automaton $(Q,A,\delta,Q_0,Q_\infty)$,
where $Q_0$ is the set of initial states and $Q_\infty = \{q_1,\ldots,q_k\}$ 
is the set of states visited infinitely often. 
On the other hand, rationality over $A$ in $(P(A^+),P(A^\omega),^+,^\omega)$ 
corresponds to the classic notion of regularity, so that $L \subseteq A^+$ 
($U \subseteq A^\omega$, resp.)  is rational over $A$ in $(P(A^+),P(A^\omega),^+,^\omega)$ 
iff it is regular. 
\end{expl}

The main result for automata is the following Kleene theorem 
that can be easily derived from the corresponding result
for partial Conway semiring-semimodule pairs \cite{Esikpartial}.

\begin{thm}
\label{thm-Kleene}
Suppose that $(H,V,^+,^\omega)$ is a Conway hemiring-hemimodule 
pair and $S_0$ is a semiring with a bi-action on $H$ and a left 
action on $V$, subject to the compatibility conditions (\ref{eq-action}),
(\ref{eq-action2}), (\ref{eq-I}), (\ref{eq-II}) of the 
previous sections. Suppose that $A\subseteq H$. 
An element of $H$ is $S_0$-rational over $A\subseteq H$ iff it is the finitary behavior of
an $S_0$-automaton over $A$. Moreover, an element of $V$ is $S_0$-rational over $A$ 
iff it is the infinitary behavior of an $S_0$-automaton over $A$. 
\end{thm}

{\sl Proof.} By Proposition~\ref{prop-sum-Conway-pair}, $(S_0 \os H,H,V,^*,^\omega)$ is a 
partial Conway semiring-semimodule pair, and $S_0$ may be identified with 
a subsemiring of $S_0 \os H$. An $S_0$-automaton over $A$ in the
Conway hemiring-hemimodule pair $(S_0 \oplus H,V,^*,^\omega)$
is then the same as an $S_0$-automaton \cite{Esikpartial} over $A$ in the partial Conway-semiring
semimodule pair $(S_0 \os H,H,V,^*,^\omega)$.

Suppose that $\cA = (\alpha,M,\beta,k)$ is such an automaton. 
Then the infinitary behavior of $\cA$ in $(H,V,^+,^\omega)$  
defined above is the same as its infinitary behavior of $\cA$ in 
the partial Conway semiring-semimodule pair $(S_0 \os H,H,V,^*,^\omega)$.
The finitary behaviors are slightly different, as in the partial Conway semiring-semimodule
pair  $(S_0 \os H,H,V,^*,^\omega)$, it is defined as $\alpha M^* \beta$ instead of $\alpha M^+ \beta$. 
But it is clear that $s \in S_0 \oplus H$  is the finitary behavior 
of an automaton over $A$ in $(S_0 \os H,H,V,^*,^\omega)$ iff $s$ can be written in the 
form $x + a$, where $x \in S_0$ and $a$ is the finitary behavior of an automaton over 
$A$ in $(H,V,^+,^\omega)$.

Regarding rational elements, we have a similar situation. When $v \in V$, then $v$ is 
rational over $A$ in $(H,V,^+,^\omega)$ iff $v$ is rational over $A$ in 
$(S_0 \os H,H,V,^*,^\omega)$, as defined in \cite{Esikpartial}. Moreover, 
$s \in S_0 \oplus H$ is rational over $A$ in $(S_0 \oplus H,H,V,^*,^\omega)$ iff it is 
of the form $x + a$ such that  $x \in S_0$ and $a\in H$ is rational over $A$ 
in $(H,V,^+,^\omega)$. Thus Theorem~\ref{thm-Kleene} follows from the Kleene 
theorem of \cite{Esikpartial}. \eop

\begin{expl}
Consider Example~\ref{expl-buchi}. Then Theorem~\ref{thm-Kleene} asserts that a language of 
finite nonempty words or a language of $\omega$-words can be accepted by an 
nfa or a B\"uchi-automaton iff it is regular. 
\end{expl} 

More examples are considered in Section~\ref{sec-revisited}.

\section{Multi-hemirings and valuation monoids}

In this section, we introduce and investigate abstract algebraic
structures which capture essential properties of the average
or discounting operations on the real numbers. As a consequence, in Section 14 we will obtain
a Kleene-type characterization of the possible behaviors of
appropriate automata models on infinite words and over such weight structures by
$\omega$-rational series. For closely related algebraic structures, we refer the reader to \cite{DrosteMeineckeMFCS, DrosteMeinecke}; also cf. \cite{Meinecke}.

\begin{deff}
A structure $(D, +, \circ, 0)$ is a {\rm multi-hemiring} if $(D, +, 0)$ is a commutative monoid and
$\circ = (\cdot_{m,n} \; | \; m,n \ge 1)$ is a family of product operations
$\cdot_{m,n}: D \times D \to D$ such that for all $a,b,c \in D$ and $k,m,n \ge 1$:
\begin{gather}
\label{eq:mh_1}
0 \cdot_{m,n} a = a \cdot_{m,n} 0 = 0 \\
\label{eq:mh_2}
(a \cdot_{k,m} b) \cdot_{k+m,n} c = a \cdot_{k, m+n} (b \cdot_{m,n} c) \\
\label{eq:mh_3}
a \cdot_{m,n} (b + c) = a \cdot_{m,n} b + a \cdot_{m,n} c \; \text{ and } \; (a + b) \cdot_{m,n} c  = a \cdot_{m,n} c + b \cdot_{m,n} c
\end{gather}
\end{deff}

Observe that equation (\ref{eq:mh_2}) is a form of associativity and equation (\ref{eq:mh_3}) is the usual distributivity law. Clearly, if $(D, +, \cdot, 0)$ is a hemiring and we put $\cdot_{m,n} = \cdot$ for all $m,n \ge 1$, then $(D, +, \circ, 0)$ is a multi-hemiring. The following examples will be important for calculating averages or discounting of weights.

\begin{example}
\label{Example:5.2}
Consider $\langle \mathbb R \cup \{-\infty\}, \sup, \circ, -\infty \rangle$ with 
$\circ = (\cdot_{m,n} \; | \; m,n \ge 1)$, given as follows.
\begin{enumerate}
\item [(a)] For $m,n \ge 1$ and $x,y \in \mathbb R \cup \{-\infty\}$, let
$x \cdot_{m,n} y = \frac{m \cdot x + n \cdot y}{m + n}$.
\item [(b)] Let $\lambda > 0$. For $m,n \ge 1$ and $x,y \in \mathbb R \cup \{-\infty\}$, let
$x \cdot_{m,n} y = x + \lambda^m \cdot y$. 
\end{enumerate}
In both cases, as usual the product is $-\infty$ if $x = -\infty$ or $y = -\infty$. Then
$(\mathbb R \cup \{-\infty\}, \sup, \circ, -\infty)$ is a multi-hemiring.
\end{example}

Next we recall from \cite{DrosteMeinecke} an abstract model
for the calculation of weights which will be used in a corresponding
weighted automaton model.

\begin{deff}
A valuation monoid $(D, +, \val, 0)$ consists of a commutative monoid $(D, +, 0)$ and a valuation function $\val: D^+ \to D$ with $\val(d) = d$ and $\val(d_1, ..., d_n) = 0$ whenever $d_i = 0$ for some $i \in \{1, ..., n\}$.
\end{deff}

The valuation function $\val$ can be seen as a very general product operation with almost no requirements (like e.g. associativity or distributivity); it incorporates classical products, but also average and discounting. To see this, next we describe the relationship between multi-hemirings and valuation monoids. Let $(D, +, \circ, 0)$ be a multi-hemiring. We define the induced valuation function $\val: D^+ \to D$ inductively by letting
\begin{eqnarray*}
\val(d) & = & d, \\
\val(d_1, ..., d_{n+1}) & = & d_1 \cdot_{1,n} \val(d_2, ..., d_{n+1})
\end{eqnarray*}
for all $d, d_1, ..., d_{n+1} \in D$ and $n \ge 1$.

Due to the identity (\ref{eq:mh_2}), it follows easily by induction that then $\val$ satisfies the following equation for all $d_i, d_j' \in D$ and $m,n \ge 1$:
$$\val(d_1, ..., d_m, d_1', ..., d_n') = \val(d_1, ..., d_m) \cdot_{m, n} \val (d_1', ..., d_n')$$

Clearly, if $(D, +, \cdot, 0)$ is a hemiring, $\circ = (\cdot_{m,n} \; | \; m,n \ge 1)$ with $\cdot_{m,n} = \cdot$ for all $m,n \ge 1$, and val is the induced valuation function, then ${\val(d_1, ..., d_n) = d_1 \cdot ... \cdot d_n}$, the usual product.

\begin{example}
\label{Example:5.12}
Consider the two multi-hemirings $(\mathbb R \cup \{-\infty\}, \sup, \circ, -\infty)$ of Example \ref{Example:5.2}, and let val be the induced valuation function.
\begin{enumerate}
\item [(a)] In case of Example \ref{Example:5.2} (a), we obtain $\val = \avg$ with
$\avg(d_1, ..., d_n) = \frac{1}{n} \sum\limits_{i = 1}^n d_i$.
\item [(b)] Let $\lambda > 0$ as in Example \ref{Example:5.2}(b). Then $\val = \disc_{\lambda}$ with $\disc_{\lambda} (d_0, ..., d_n) = \sum\limits_{i = 0}^n \lambda^i d_i$.
\end{enumerate}
\end{example} 
Now we turn to valuations of infinite sequences of weight. A monoid $(D, +, 0)$ is {\em complete}, cf. \cite{Eilenberg}, if it has infinitary sum operations $\sum_{I}: D^I \to D$ for any index set $I$ such that
\begin{itemize}
\item $\sum_{i \in \emptyset} d_i = 0$, $\sum_{i \in \{k\}} d_i = d_k$, $\sum_{i \in \{j,k\}} d_i = d_j + d_k$ for $j \neq k$, and
\item $\sum_{j \in J} \big( \sum_{i \in I_j} d_i \big) = \sum_{i \in I} d_i$ if $\bigcup_{j \in J} I_j = I$ and $I_j \cap I_k = \emptyset$ for $j \neq k$.
\end{itemize}
\begin{deff}
A structure $(D, +, (\cdot_{m,n} \; | \; m \in \N, n \in \mathbb N \cup \{\omega\}), \val^{\omega}, 0)$ is an {\rm $\omega$-valuation multi-hemiring}, if 
\begin{itemize}
\item $(D, +, (\cdot_{m,n} \; | \; m,n \in \N), 0)$ is a multi-hemiring,
\item $(D, +, 0)$ is a complete monoid,
\item $\cdot_{m, \omega}: D \times D \to D$ for every $m \in \mathbb N$, and
$\val^{\omega}: (\N \times D)^{\omega} \to D$ such that $\val^{\omega}(n_i, d_i)_{i \ge 1} = 0$  whenever $d_i = 0$ for some $i \ge 1$,
\item the following equalities hold for all $d,d',d''$ and $d_i,d_j'$ in $D$,
where $i \geq 1$ and $j \in I$ and $I$ is an arbitrary index set:
\begin{gather}
\label{eq:omh_pr1}
0 \cdot_{m, \omega} d = d \cdot_{m, \omega} 0 = 0 \\
\label{eq:omh_pr2}
d \cdot_{m, \omega} (d' \cdot_{n, \omega} d'') = (d \cdot_{m, n} d') \cdot_{m+n, \omega} d'' \\
\label{eq:omh_pr3}
d \cdot_{m, \omega} \sum\limits_{i \in I} d_i' = \sum\limits_{i \in I} d \cdot_{m, \omega} d_i' \; \text{ and } \; \bigg(\sum_{i \in I} d_i' \bigg) \cdot_{m,\omega} d = \sum_{i \in I} d_i' \cdot_{m, \omega} d  \\
\label{eq:omh_pr4}
\val^{\omega} (n_i, d_i)_{i \ge 1} = d_1 \cdot_{n_1, \omega} \val^{\omega} (n_i, d_i)_{i \ge 2} 
\end{gather}
for all $n_k \in \N$, $d_{i_k} \in D$ ($i_k \in I_k$) with finite index sets $I_k$ ($k \ge 1$):
\begin{equation}
\label{eq:omh_pr5}
\val^{\omega}\left(n_k, \sum\limits_{i_k \in I_k} d_{i_k} \right)_{k \ge 1} =
\sum\limits_{(i_k)_k \in I_1 \times I_2 \times ... } \val^{\omega}(n_k, d_{i_k})_{k \ge 1}
\end{equation}
\end{itemize}
\end{deff}
We say that the  $\omega$-valuation multi-hemiring is {\it infinitary associative}, if it satisfies for all $m_{i} \ge 1$ and $d_i \in D$ ($i \ge 1$) and each subsequence $(n_i)_{i \ge 1}$ of $\mathbb N$ the equation
\begin{eqnarray}
\label{eq:omh_pr6}
\val^\omega (k_j, d'_j) &=& \val^\omega (m_i, d_i)
\end{eqnarray}
where 
\begin{eqnarray*}
k_j  &=& m_{n_{j-1} + 1} + ... + m_{n_j}\\
d'_j &=& (...(d_{{n_{j-1} + 1}} \cdot_{m_{n_{j-1} + 1}, m_{n_{j-1} + 2}} d_{{n_{j-1} + 2}}) \cdot ...  \notag \\
&& \cdot d_{{n_j - 1}}) \cdot_{m_{n_{j-1} + 1} + ... + m_{n_{j} - 1}, m_{n_j}} d_{{n_j}} 
\end{eqnarray*}
for each $j \geq 1$ (letting $n_0 = 0$).

These conditions are similar to (and slightly stronger than) corresponding ones of  \cite{DrosteMeinecke} and  \cite{Meinecke} for Cauchy $\omega$-indexed valuation monoids. For the interpretation of these conditions, it is useful to consider $\val^{\omega}$ as a parameterized (by $\omega$-sequences over $\N$) infinitary product on $D$. Properties (\ref{eq:omh_pr2}) and (\ref{eq:omh_pr4}) are a form of finitary associativity. Property (\ref{eq:omh_pr3}) is distributivity of the multiplication $\cdot_{m, \omega}$, and property (\ref{eq:omh_pr5}) is an infinitary distributivity of $\val^{\omega}$. Also, condition (\ref{eq:omh_pr2}) is slightly stronger than condition (13) of
 \cite{Meinecke} (where, in comparison, $d''$ is of the specific form $\val^{\omega}(n_i, d_i)_{i \ge 3}$).

Infinitary associative $\omega$-valuation multi-hemirings are related to the 
complete $\omega$-hemirings equipped with an infinitary product. The definition 
below is motivated by the notion of complete semirings \cite{Eilenberg} and the 
complete semirings of \cite{EsikKuichsem1,EsikKuichsem2} equipped with an infinitary product.

\begin{deff}
A \emph{complete $\omega$-hemiring} is a hemiring $H$ such that $(H, +, 0)$ is a complete monoid with infinitary sums $\sum_I$ and $H$ is equipped with an infinitary product operation $(a_1,a_2,\ldots) \mapsto \prod_{i \geq 1}a_i$
mapping infinite sequences of elements of $H$ to $H$, subject to the following 
axioms:
\begin{eqnarray}
b(\sum_{i \in I}a_i) &=& \sum_{i \in I} ba_i \nonumber\\
(\sum_{i \in I} a_i)b &=& \sum_{i \in I} a_ib \nonumber \\
\label{eq-infprod}
a_1 \prod_{j \geq 2}a_j &=& \prod_{j\geq 1}a_j\\
\label{eq_52}
\prod_{j \geq 1}a_j
&=& 
\bigg(\prod_{0 <  j \leq i_1} a_j \bigg) \cdot \bigg( \prod_{i_1 < j \leq i_2}a_j \bigg)\cdot \ldots \\
\prod_{j \geq 1} \sum_{i \in I_j}a_i &=& 
\sum_{(i_1,i_2,\ldots) \in I_1 \times I_2 \times ...}\prod_{j\geq 1}a_{i_j}
\end{eqnarray}
where in equation (\ref{eq_52}), $i_1 < i_2 < \ldots$ is an arbitrary sequence of positive integers. 
\end{deff}

Using methods from \cite{EsikKuichsem1}, it is easy to show that every complete $\omega$-hemiring $H$
gives rise to an iteration hemiring-hemimodule pair $(H,H,^+,^\omega)$, where 
$a^+ = \sum_{i\geq 1}a^i$ and $a^\omega = \prod_{i \geq 1} a = aa\cdots$.
Also, every complete $\omega$-hemiring is an infinitary associative $\omega$-valuation multi-hemiring 
with $a \cdot_{m,n} b = ab$ and $\val^\omega ((n_1,a_1),(n_2,a_2),\ldots) = \prod_{i \geq 1}a_i$.

Now we give several examples of $\omega$-valuation multi-hemirings. The calculations can be done 
very similarly to the ones of \cite{DrosteMeinecke}; this is left to the reader.

\begin{expl}
Let $L = (L,\vee,\wedge,0,1)$ be a completely distributive lattice. Then 
$(L,\vee,\wedge,\inf,0)$ is an infinitary associative $\omega$-valuation
multi-hemiring which can be derived from a complete $\omega$-hemiring. 
\end{expl} 

Next, we investigate five examples of weight structures considered in \cite{Chatterjeeetal, Chatterjeeetal2} and show how they fit into our framework. We replace their value functions on $\mathbb Q^{\omega}$ by functions on $\oR_+^{\omega}$ where  \linebreak $\oR_+ = \{x \in \R : x \geq 0\} \cup \{-\infty,\infty\}$ in order to avoid convergence issues for infinite sums. Example \ref{expl-three} (b),(d),(e) are essentially due to \cite{DrosteMeinecke}; we repeat the argument for the sake of completeness.

\begin{expl}
\label{expl-three}
Consider $D = (\oR_+,\sup, (\cdot_{m,n} : m \in \N,n \in \N \cup \{\omega\}), \val^\omega, -\infty)$
where $\cdot_{m,n}$ and $\cdot_{m,\omega}$ ($m,n \in \N$) and $\val^\omega$ are given as follows, 
in each case ensuring the necessary conditions for $-\infty$ as zero.
\begin{enumerate}
\item [(a)]
\begin{itemize}
\item
$a\cdot_{m,n} b = a \cdot_{m, \omega} b = \sup\{a,b\}$ if $a,b\geq 0$;
\item
$\val^\omega(n_i,a_i)_{i \geq 1} = \sup_{i \geq 1}a_i$ if $a_i \geq 0$ for all $i\geq 1$.
\end{itemize}
Then $D$ is an infinitary associative $\omega$-valuation multi-hemiring
which can be derived as above from a complete $\omega$-hemiring having $\prod_{i \ge 1} a_i = \sup_{i \ge 1} a_i$ as product operation.
\item [(b)]
\begin{itemize}
\item
$a\cdot_{m,n} b = \sup\{a,b\}$ and $a \cdot_{m,\omega} b = b$ if $a,b\geq 0$;
\item
$\val^\omega(n_i,a_i)_{i \geq 1} = \lim\sup_{i \geq 1}a_i$ if $a_i \geq 0$ for all $i\geq 1$.
\end{itemize}
Then $D$ is an infinitary associative $\omega$-valuation multi-hemiring
which cannot be derived as above from a complete $\omega$-hemiring since (\ref{eq-infprod}) fails. Note that if here we let $a \cdot_{m, \omega} b = \sup \{a, b\}$ if $a,b \ge 0$, then equation (\ref{eq:omh_pr4}) would fail.
\item [(c)] \begin{itemize}
\item $a \cdot_{m,n} b = \sup \{a, b\}$ and $a \cdot_{m, \omega} b = b$ if $a,b \ge 0$; 
\item $\val^{\omega}(n_i, a_i) = \liminf_{i \ge 1} a_i$ if $a_i \ge 0$ for all $i \ge 1$.
\end{itemize}
Then $D$ is an $\omega$-valuation multi-hemiring, but $D$ is not infinitary associative. Indeed, let $m_i = 1$, $d_{2i} = 1$, $d_{2i-1} = 0$, and $n_i = 2i$ for all $i \ge 1$. Then
$\val^{\omega}(1, d_i) = \liminf_{i \ge 1} d_i = 0$, but $\val^{\omega}(2, \sup\{d_{2i-1}, d_{2i}\})_{i \ge 1} = \liminf_{i \ge 1} d_{2i} = 1$.
\item [(d)] Let $0 < \lambda < 1$. Put
\begin{itemize}
\item $a\cdot_{m,n} b = a\cdot_{m,\omega} b = a + \lambda^m b$ if $a,b\geq 0$;
\item $\val^\omega(n_1,a_i)_{i \geq 1} = 
\lim_{k \to \infty} (a_1 + \lambda^{n_1}a_2 + \ldots + \lambda^{n_1+\ldots + n_{k-1}}a_k)$ 
if $a_i \geq 0$ for all $i$.
\end{itemize}
Then $D$ is an infinitary associative $\omega$-valuation multi-hemiring which cannot 
be derived from a complete $\omega$-hemiring. 
\item [(e)]
\begin{itemize}
\item $a \cdot_{m,n}b = (ma + nb)/(m + n)$ and $a\cdot_{m,\omega} b = b$ for all $a,b\geq 0$;
\item $\val^\omega(n_i,a_i)_{i \geq 1} = \lim\sup\avg (n_i,a_i)_{i \geq 1} = 
     \lim\sup_k (n_1a_1 + \ldots +n_ka_k)/(n_1 + \ldots +  n_k)$ if $a_i \geq 0$ for all $i\geq 1$.
\end{itemize}
Then $D$ is an $\omega$-valuation multi-hemiring. But $D$ is not infinitary associative. This, and a slightly stronger statement, will also follow from the subsequent Example 13.10. For a direct proof, consider the $\omega$-word $w = (1, 0)^1 (1, 1)^2 (1, 0)^4 (1, 1)^8 (1, 0)^{16} ... \in (\mathbb N \times \{0,1\})^{\omega}$; i.e., in the second component sequences of $0$'s and $1$'s are alternating doubling their lengths at each alternation. Then $\lim \sup \avg(w) = \frac{2}{3}$. But 
$$\begin{array}{@{}l}
\lim \sup \avg((1, 0), (6, \avg(1^2 0^4)), (24, \avg(1^80^{16})), ...) \\
= \lim \sup \avg((1, 0), (6, \frac{1}{3}), (24, \frac{1}{3}), ...) = \frac{1}{3},
\end{array}
$$
so equation (50) is violated.

It is easy to see that this phenomenon occurs for a sequence $w' \in \{0,1\}^{\omega}$ iff $\lim \sup \avg(w') \neq \lim \inf \avg (w')$.

\end{enumerate}
\end{expl}
Next we wish to consider series over $\Sigma^{\omega}$. Let $D$ be an $\omega$-valuation multi-hemiring. We let $D \llangle \Sigma^{\omega} \rrangle$ comprise all functions ({\em series}) from $\Sigma^{\omega}$ to $D$. As usual, when $s \in D \llangle \Sigma^{\omega} \rrangle$ and $w \in \Sigma^{\omega}$, we write $(s, w)$ for $s(w)$.

For $r,s\in D \llangle \Sigma^{\omega} \rrangle$ we define $r+ s$ by $(r+s,w) = (r,w) + (s,w)$ 
for all $w \in \Sigma^{\omega}$. Now if $r,s\in D\llangle \Sigma^+\rrangle$ and $s' \in D\llangle\Sigma^\omega\rrangle$,
we define  $rs\in D\llangle \Sigma^+ \rrangle $ and $rs'\in D\llangle \Sigma^\omega \rrangle$ 
by letting
\begin{eqnarray*}
(rs,w) &=& \sum \{(r,u)\cdot_{|u|,|v|} (s,v) : u,v\in \Sigma^+,\ uv = w\}, \quad (w \in \Sigma^+)\\
(rs',w) &=& \sum \{ (r,u)\cdot_{|u|,\omega} (s',v) : u \in \Sigma^+,\ v \in \Sigma^\omega \ uv = w \}, 
\quad (w \in \Sigma^\omega).
\end{eqnarray*}
For a sequence of series $r_i\in D\llangle \Sigma^+\rrangle$ ($i \geq 1$), we define the infinite 
product $r = \prod_{i \geq 1}r_i \in D\llangle \Sigma^\omega \rrangle$ by 
\begin{eqnarray*}
(r,w) &=& \sum\{\val^\omega (|u_k|,(r,u_k)) : w = u_1u_2\ldots,\ u_k\in \Sigma^+,\ k \geq 1\}
\end{eqnarray*} 
for all $w \in \Sigma^\omega$.
We also define the \emph{powers} of $r$ by $r^1 = r$ and $r^{n + 1} = rr^n$, for all $n \geq 1$.
As usual, we define
\begin{eqnarray*}
r^+ &=& \sum_{i \geq 1}r^i\\
r^\omega &=&\prod_{i \geq 1}r
\end{eqnarray*} 
for all $r \in D\llangle \Sigma^+\rrangle$. 
Note that 
\begin{eqnarray*}
(r^+,w) &=& \sum_{i =1}^{|w|}(r^i,w),\quad (w \in \Sigma^+)\\
(r^\omega,w)  &=& \sum\{\val^\omega ( |u_k|, (r,u_k)) : w = u_1u_2\ldots, \ 
 u_k \in \Sigma^+,\ k = 1,2,\ldots\},\quad (w \in \Sigma^\omega).
\end{eqnarray*} 
(The above sums all exist since $D$ is a complete monoid). 

Now we show:

\begin{thm}
\label{thm-multi}
Let $D$ be an infinitary associative $\omega$-valuation multi-hemiring and $\Sigma$ an alphabet. Then $(D \llangle \Sigma^+ \rrangle, D \llangle\Sigma^{\omega} \rrangle)$ is a complete hemiring-\-hemimodule pair.
\end{thm}

{\sc Proof.}
By Proposition 5.3 of \cite{DrosteKuich}, $D \llangle \Sigma^+ \rrangle$ is a hemiring. Here we just note that the associativity of the Cauchy product follows from equations (\ref{eq:mh_2}) and (\ref{eq:mh_3}) and the distributivity over sum from equation (\ref{eq:mh_3}).

Clearly, since $D$ is a complete monoid, so is $(D \llangle \Sigma^{\omega}\rrangle, +, 0)$. Next we claim that $r_1(r_2s) = (r_1 r_2) s$ for any $r_1, r_2 \in D \llangle \Sigma^+ \rrangle$ and $s \in D \llangle \Sigma^{\omega} \rrangle$. Let $w \in \Sigma^{\omega}$. Then 
\begin{eqnarray*}
(r_1 (r_2 s), w) &=& \sum\limits_{w = u_1 v_1}  (r_1, u_1) \cdot_{|u_1|, \omega} (r_2s, v_1) \\
& = &  \sum\limits_{w = u_1 v_1}  (r_1, u_1) \cdot_{|u_1|, \omega} \sum\limits_{v_1 = u_2 v}
(r_2, u_2) \cdot_{|u_2|, \omega} (s, v) \\
& = & \sum\limits_{w = u_1 u_2 v}  (r_1, u_1) \cdot_{|u_1|, \omega} ((r_2, u_2) \cdot_{|u_2|, \omega} (s, v)), \\
((r_1 r_2) s), w) &=& \sum\limits_{w = u v}  (r_1 r_2, u) \cdot_{|u|, \omega} (s, v) \\
& = & \sum\limits_{w = u v}  \left( \sum\limits_{u = u_1 u_2} (r_1, u_1) \cdot_{|u_1|, |u_2|}
(r_2, u_2) \right) \cdot_{|u|, \omega} (s, v) \\
& = &  \sum\limits_{w = u_1 u_2 v}  ((r_1, u_1) \cdot_{|u_1|, |u_2|} (r_2, u_2)) \cdot_{|u_1|+|u_2|, \omega} (s, v).
\end{eqnarray*}
Now equation (46) implies our claim.

Next, let $r_i \in D \llangle \Sigma^+ \rrangle$ ($i \ge 1$). We show that
$\prod_{i \ge 1} r_i = r_1 \cdot \prod_{i \ge 2} r_i$. Indeed, let $w \in \Sigma^{\omega}$. Then
\begin{eqnarray*}
\left( r_1 \cdot \prod\nolimits_{i \ge 2} (r_i, w) \right) &=& \sum_{w = u_1 w_1} (r_1, u_1) \cdot_{|u_1|, \omega} \sum_{w_1 = u_2 u_3 ...} \val^{\omega} (|u_i|, (r_i, u_i))_{i \ge 2} \\
&=& \sum_{w = u_1 u_2 u_3 ... } (r_1, u_1) \cdot_{|u_1|, \omega} \val^{\omega}(|u_i|, (r_i, u_i))_{i \ge 2} \\
&\stackrel{(\ref{eq:omh_pr4})}{=}& \sum_{w = u_1 u_2 u_3 ... } \val^{\omega}(|u_i|,
(r_i, u_i))_{i \ge 1} \\
&=& \left( \prod\nolimits_{i \ge 1} r_i, w \right).
\end{eqnarray*}
Next, let $r_{i_j} \in D \llangle \Sigma^+ \rrangle$ for $i_j \in I_j$ with index sets $I_j$, for $j \ge 1$. We wish to show the infinitary distributivity law
$\prod\limits_{j \ge 1} \left( \sum\limits_{i_j \in I_j} r_{i_j} \right) = 
\sum\limits_{(i_k)_k \in I_1 \times I_2 \times ... } \; \prod\limits_{j \ge 1} r_{i_j}$. Let $w \in \Sigma^{\omega}$. Then
\begin{eqnarray*}
\left( \left( \sum_{(i_k)_k \in I_1 \times I_2 \times ...} \; \prod_{j \ge 1} r_{i_j} \right), w) \right)
&=& \sum_{(i_k)_k \in I_1 \times I_2 \times ... } \left( \prod_{j \ge 1} r_{i_j}, w \right) \\
&=& \sum_{(i_k)_k \in I_1 \times I_2 \times ... } \; \sum_{w = u_1 u_2 ...}
\val^{\omega}(|u_j|, (r_{i_j}, u_j))_{j \ge 1} \\
&=& \sum_{w = u_1 u_2 ...} \; \sum_{(i_k)_k \in I_1 \times I_2 \times ... } \val^{\omega}(|u_j|, (r_{i_j}, u_j))_{j \ge 1} \\
&\stackrel{(\ref{eq:omh_pr5})}{=}& \sum_{w = u_1 u_2 ...} \val^{\omega} \left(|u_j|,
\sum\nolimits_{i_j \in I_j} (r_{i_j}, u_j)\right)_{j \ge 1} \\
&=& \left(\prod_{j \ge 1} \left(\sum_{i_j \in I_j} r_{i_j} \right), w \right).
\end{eqnarray*}
Finally, let $r_i \in D \llangle \Sigma^+ \rrangle$ for $i \ge 1$, and let
$0 = n_0 < n_1 < n_2 ...$ in $\N$. We wish to show the infinitary associativity law that
$\prod_{i \ge 1} r_i = \prod_{j \ge 1} (r_{n_{j-1}+1} \cdot ... \cdot r_{n_j})$. Let $w \in \Sigma^{\omega}$. Then
\begin{eqnarray*}
 \lefteqn{\left(\prod\nolimits_{j \ge 1} (r_{n_{j-1} +1} \cdot ... \cdot r_{n_j}, w \right)} \\
&=& \sum_{w = u_1 u_2 ...} \val^{\omega}(|u_j|, (r_{n_{j-1} + 1} \cdot ... \cdot r_{n_j}, u_j))_{j \ge 1} \\
&=& \sum_{w = u_1 u_2 ...} \val^{\omega} \Biggl(|u_j|, \sum_{u_j = v_{n_{j-1}+1} ... v_{n_j}} (...(((r_{n_{j-1}+1}, v_{n_{j-1} + 1}) \cdot_{|v_{n_{j-1}+1}|, |v_{n_{j-1} + 2}|}  \\ 
&& (r_{n_{j-1} + 2}, v_{n_{j-1} + 2})) \cdot_{|v_{n_{j-1} + 1}| + |v_{n_{j-1} + 2}|, |v_{n_{j-1} + 3}|}
(r_{n_{j-1} + 3}, v_{n_{j-1} +3}))  \\ 
&& \cdot ... \cdot (r_{n_j - 1}, v_{n_j - 1})) \cdot_{|v_{n_{j-1} + 1}| + ... + |v_{n_j -1}|, |v_{n_j}|} (r_{n_j}, v_{n_j}) \Biggr)_{j \ge 1} \\
& \overset{(49)}{=} & \sum_{w = u_1 u_2 ... } \; \sum_{\begin{subarray}{c} u_j = v_{n_{j-1} + 1} ... v_{n_j} \\ (j \ge 1) \end{subarray}} \val^{\omega} \left(|u_j|, (...((r_{n_{j-1} + 1}, v_{n_{j-1} + 1} ) \cdot_{|v_{n_{j-1} + 1}|, |v_{n_{j-1} + 2}|} \right.  \\
&& \left. (r_{n_{j-1}+2}, v_{n_{j-1} + 2})) \cdot ... \cdot (r_{n_j - 1}, v_{n_j - 1})) 
\cdot_{|v_{n_{j-1} + 1}| + ... + |v_{n_j -1}|, |v_{n_j}|} (r_{n_j}, v_{n_j}) \right)_{j \ge 1} \\
& = & \sum_{w = v_1 v_2 ... } \val^{\omega} (|v_i|, (r_i, v_i))_{i \ge 1}
\end{eqnarray*}
using infinitary associativity (\ref{eq:omh_pr6}). The latter sum equals $\left(\prod_{i \ge 1} r_i, w \right)$, proving our claim.
\eop

As an immediate consequence of Theorem~\ref{thm-multi} and results 
in \cite{EsikKuichsem1}, if $D$ is an infinitary associative
$\omega$-valuation multi-hemiring and $\Sigma$ is an alphabet, 
then $(D\llangle \Sigma^+\rrangle,D\llangle \Sigma^\omega\rrangle)$ is an 
iteration semiring-semimodule pair. In particular this applies 
to Example~\ref{expl-three} (a) and (b). This is contrasted by the 
$\omega$-valuation multi-hemiring of Example~\ref{expl-three} (c) as we show now.

\begin{example}
Consider the $\omega$-valuation multihemiring $D = (\oR_+,\sup, (\avg_{m,n}:
m\in \N,n \in \N \cup \{\omega\}), \lim\sup\avg, -\infty)$ of Example~\ref{expl-three}
(c). Let $\Sigma$ be an alphabet. We claim that $(D\llangle \Sigma^+\rrangle,
D\llangle \Sigma^\omega \rrangle,^+,^\omega)$ with the plus and omega power 
operations defined as above does not satisfy the product 
omega identity, so that is is not Conway. 
Consequently, by Theorem~\ref{thm-multi}, $D$ is not infinitary associative (which
we already saw directly in Example 13.8(c)). 

Now let $r,s \in D\llangle \Sigma^+\rrangle$ and $w\in \Sigma^\omega$.
To consider the product omega identity $(r \cdot s)^{\omega} = r \cdot (s \cdot r)^{\omega}$, we calculate in general:

\begin{eqnarray*}
\left( \left(r \cdot s\right)^{\omega}, w \right) &=& \sum_{w = w_1 w_2 ... } 
\val^{\omega} (|w_i|, (r \cdot s, w_i))_{i \ge 1} \\
& = & \sum_{w = w_1 w_2 ... } \val^{\omega} \left(|w_i|, \sum_{w_i = u_i v_i}
(r, u_i) \cdot_{|u_i|, |v_i|} (s, v_i) \right)_{i \ge 1} \\
& \stackrel{(\ref{eq:omh_pr5})}{=} & \sum_{w = w_1 w_2 ...} 
\; \sum_{\begin{subarray}{c} w_i = u_i v_i \\ (i \ge 1) \end{subarray}} 
\val^{\omega} \left(|w_i|, (r, u_i) \cdot_{|u_i|, |v_i|} (s, v_i) \right)_{i \ge 1} \\
& = & \sum_{w = u_1 v_1 u_2 v_2 ... } \val^{\omega} \left( |u_i| + |v_i|,
(r, u_i) \cdot_{|u_i|, |v_i|} (s, v_i) \right)_{i \ge 1}
\end{eqnarray*}

\begin{eqnarray*}
\lefteqn{(r \cdot (s \cdot r)^{\omega}, w)} \\
&=& \sum_{w = u w'} (r, u) \cdot_{|u|, \omega}
\left( \left(s \cdot r \right)^{\omega}, w' \right) \\
& = & \sum_{w = uw'} (r, u) \cdot_{|u|, \omega} \sum_{w' = w_1' w_2' ...} 
\val^{\omega} \left( |w_i'|, \sum_{w_i' = v_i' u_i'} (s, v_i') \cdot_{|v_i'|, |u_i'|} (r, u_i') \right)_{i \ge 1} \\
& \stackrel{(\ref{eq:omh_pr5})}{=} & 
\sum_{w = uw'} (r, u) \cdot_{|u|, \omega} \sum_{w' = w_1' w_2' ...} 
\sum_{\begin{subarray}{c} w_i' = v_i' u_i' \\ (i \ge 1) \end{subarray}}
\val^{\omega} \left( |w_i'|, (s, v_i') \cdot_{|v_i'|, |u_i'|} (r, u_i') \right)_{i \ge 1} \\
& \stackrel{\begin{subarray}{l} (\ref{eq:omh_pr3}), \\ (\ref{eq:omh_pr4}) \end{subarray}}{=} &
\sum_{w = uv_1'u_1'v_2'u_2'...} \val^{\omega} \left((|u|, (r, u)), \left(|v_i' + u_i'|, 
(s, v_i') \cdot_{|v_i'|, |u_i'|} (r, u_i') \right)_{i \ge 1} \right) 
\end{eqnarray*}

{\it Equality} would be obtained if
\begin{align*}
& \val^{\omega} \left( |u_i| + |v_i|,
(r, u_i) \cdot_{|u_i|, |v_i|} (s, v_i) \right)_{i \ge 1} \\
& = \val^{\omega} \left((|u_1|, (r, u_1)), \left(|v_i| + |u_{i+1}|, 
(s, v_i) \cdot_{|v_i|, |u_{i+1}|} (r, u_{i+1}) \right)_{i \ge 1} \right) 
\end{align*}

To obtain a counterexample,
choose a sequence $n_1 < n_2 < n_2 < ...$ in $\mathbb N$ which is quickly ascending so that
$\frac{1}{n_{k+1}} \cdot \sum_{i = 1}^k n_i \to 0$ for $k \to \infty$. Let $u_i = a^{n_i}, 
v_i = b^{n_i} \in \Sigma^*$ ($i \ge 1$), and define $r,s \in \overline{\mathbb R_+} \llangle \Sigma^* \rrangle$ with $\supp(r) = \{u_i \; | \; i \ge 1\}$, $\supp(s) = \{v_i \; | \; i \ge 1\}$, $(r, u_i) = 1$, $(s, v_i) = 0$ for $i \ge 1$. 
Next we use the notation 
$\avg((n_1, d_1), ..., (n_k, d_k)) = \frac{n_1 \cdot d_1 + ... + n_k \cdot d_k}{n_1 + ... + n_k}$. Then
$$\avg(|u_i + v_i|, (r, u_i) \cdot_{|u_i|, |v_i|} (s, v_i))_{i = 1}^k = \avg \left(2n_i, \frac{n_i + 0}{2n_i} \right)_{i = 1}^k = \frac{1}{2}$$ 
for each $k \ge 1$, so for $w = u_1 v_1 u_2 v_2 ... \in \Sigma^{\omega}$ we obtain
$$((r \cdot s)^{\omega}, w) = \text{lim\:sup\:avg}(|u_i| + |v_i|, (r, u_i) \cdot_{|u_i|, |v_i|} (s, v_i))_{i \ge 1} = \frac{1}{2}.$$
But 
\begin{gather*}
\avg((|u_1|, (r, u_1)), (|v_i| + |u_{i+1}|, (s, v_i) \cdot_{|v_i|, |u_{i+1}|} (r, u_{i+1}))_{i = 1}^k \\
= \frac{\sum_{i = 1}^{k+1} |u_i| \cdot (r, u_i)}{\sum_{i = 1}^k(|u_i| + |v_i|) + u_{k+1}} = 
\frac{\sum_{i = 1}^k n_i + n_{k+1}}{2 \cdot \sum_{i = 1}^k n_i + n_{k+1}} \to 1
\end{gather*}
for $k \to \infty$. Hence
$$(r \cdot (sr)^{\omega}, w) = \text{lim\:sup\:avg} ((|u_1|, (r, u_1)),
(|v_i| + |u_{i+1}|, (s, v_i) \cdot_{|v_i|, |u_{i+1}|} (r, u_{i+1}))_{i \ge 1}) = 1$$
proving that $(rs)^{\omega} \neq r \cdot (s \cdot r)^{\omega}$.
\end{example}

\section{Automata, revisited}
\label{sec-revisited}

In this section, we will apply our previous results to obtain a Kleene-type result for weighted automata over $\omega$-valuation multi-hemirings and infinite words. By Example \ref{expl-three}, this applies in particular to some of the weight structures investigated in \cite{Chatterjeeetal, Chatterjeeetal2}.

Let $D$ be an $\omega$-valuation multi-hemiring and $Z$ any set. 
We may think of $Z$ as the set $\Sigma^+$ of all finite non-empty 
words or the set $\Sigma^\omega$ of all $\omega$-words over the 
alphabet $\Sigma$. For $s \in D \llangle Z\rrangle$, let $\supp(s) = \{z \in Z : (s,z)\neq 0\}$,
the \emph{support} of $s$. We call $s$ a {\em polynomial} when $\supp(s)$ is finite,
and a \emph{monomial} if $\supp(s)$ is a singleton or empty. We let $D\langle \Sigma \rangle$ 
denote the set of all polynomials in $D\llangle \Sigma^+ \rrangle$ whose support is a subset 
of $\Sigma$.

Now a series $s \in D\llangle \Sigma^+\rrangle$ is \emph{rational} if it is $\N$-rational 
over $D\langle \Sigma \rangle$ as defined in Section~\ref{sec-automata}, in other words, 
when $s$ can be generated
from the monomials by the rational operations sum, product and plus. Moreover, a series 
$s \in D\llangle \Sigma^\omega \rrangle$ is \emph{$\omega$-rational} if $s$ can be generated from
the monomials by the above operations and omega power. 

An \emph{automaton} $\mathfrak{A} = (n,I,\gamma,F,k)$ over $D$ and $\Sigma$ is given by 
\begin{itemize}
\item a finite nonempty set of states $Q = \{1,\ldots,n\}$,
\item subsets $I$ and $F$ of initial and final states,
\item a set $\{1,\ldots,k\}$ of repeated states,
\item a weight function $\gamma : Q \times \Sigma \times Q \to D$.
\end{itemize}
A \emph{finite run} of $\mathfrak{A}$ is a sequence $R = (t_i)_{1 \leq i \leq n}$ of matching
transitions $t_i = (q_{i-1},a_i,q_i)$. Then $R$ is a run on $a_1\ldots a_n$. A finite run $R$ 
is \emph{successful} if it starts in an initial state from $I$ and ends in a final state from $F$. 
Moreover, $\gamma(R) = (\gamma(t_i))_{1 \leq i \leq n}$ is the sequence of weights of $R$, 
and $\wt(R) = \val(\gamma(R))$ is the \emph{weight} of $R$.

Similarly, a sequence $R = (t_i)_{i \in \N}$ of matching transitions 
$(q_{i-1},a_i,q_i)$ is an infinite run on $w= a_1a_2 \ldots$. It is 
successful if it starts in an initial state from $I$ and infinitely often passes through the 
set of repeated states $\{1,\ldots,k\}$.  We put $\gamma(R) = (\gamma(t_i))_{i \in \N}$ 
and $\wt(R) = \val^\omega(\gamma(R))$ the \emph{weight} of $R$.

The {\em finitary behavior} of $\mathfrak{A}$ is the series $|\mathfrak{A}|_f \in D\llangle \Sigma^+\rrangle$,
defined for $w \in \Sigma^+$ by
$$ (|\mathfrak{A}|_f,w) = \sum \{ \wt(R) : R\ {\rm is}\ {\rm a}\ {\rm successful}\ {\rm run}\ {\rm on}\ w \}.$$
The \emph{infinitary behavior} $|\mathfrak{A}|_\omega \in D\llangle \Sigma^\omega \rrangle$ 
of $\mathfrak{A}$ is defined in the same way:
$$ (|\mathfrak{A}|_\omega,w) = \sum \{\wt(R) : R\ {\rm is}\ {\rm a}\ {\rm successful}\ {\rm run}\ {\rm on}\ w\}.$$
 
Now if $D$ is infinitary associative, as noted before, $(D\llangle \Sigma^+ \rrangle, 
D\llangle \Sigma^\omega \rrangle,^+,^\omega)$ is a Conway hemiring-hemimodule pair.
Consider the semiring $\N$ of natural numbers with the natural bi-action on $D\llangle \Sigma^+\rrangle$ 
and left action on $D\llangle \Sigma^\omega \rrangle$. Each automaton $\mathfrak{A} = 
(n,I,\gamma,F,k)$ corresponds to an $\N$-automaton $\mathcal{A} =
(\alpha,M,\beta,k)$ over $D\langle \Sigma \rangle$ (cf. Section~\ref{sec-automata}), where $\alpha$ and $\beta$ 
take on as values only $0$ and $1$ (corresponding to membership in $I$ and $F$, respectively), 
and the matrix $M \in D\langle \Sigma \rangle^{n \times n}$ has as $(i,j)$th entry the sum of monomials
$\sum_{a \in \Sigma}\gamma(i,a,j)a$. Arguing as in \cite{EsikKuichsem1}, p. 251, we obtain that 
$|\mathfrak{A}|_f = |\mathcal{A}|_f$ and $|\mathfrak{A}|_\omega = |\mathcal{A}|_\omega$.
   
Conversely, every $\N$-automaton $\mathcal{A} = (\alpha,M,\beta,k)$ can be replaced by a disjoint union 
of such automata whose initial and final vectors only assume the values  $0$ and $1$, such that the 
disjoint union has the same finitary and infinitary behavior as $\mathcal{A}$. Then such an automaton 
corresponds to an automaton over $D$ and $\Sigma$ with the same finitary and infinitary behavior.

From Theorems \ref{thm-Kleene} and \ref{thm-multi} we immediately have: 

\begin{cor}
Let $D$ be an infinitary associative $\omega$-valuation multi-hemiring and $\Sigma$ 
an alphabet. Then a series of $D\llangle \Sigma^+\rrangle$ is rational iff it is the 
finitary behavior of an automaton over $D$ and $\Sigma$.  
Moreover, a series of $D\llangle \Sigma^\omega\rrangle$ is $\omega$-rational iff it is the 
infinitary behavior of an automaton over $D$ and $\Sigma$.
\end{cor}

This result applies in particular to the weight structures of Example \ref{expl-three} (a),(b),(d). In \cite{DrosteMeinecke} a Kleene-type result was given also for the average weight setting of Example \ref{expl-three} (e), using additional properties of the involved $\omega$-valuation multi-hemiring $D$. It remains an open challenge whether this average weight multi-hemiring can be covered by suitable Conway axioms similar to those investigated here.

\end{document}